\documentclass[superscriptaddress]{revtex4}
\usepackage{graphicx,epsfig}

\begin{document}

\newcommand{\dd} {\mbox{d\raisebox{0.75ex}{\hspace*{-0.32em}-}\hspace*{-0.02em}}}
\newcommand{\DD} {\mbox{D\raisebox{0.30ex}{\hspace*{-0.75em}-}\hspace*{ 0.42em}}}

\title{Differential cross sections of the charge-exchange reaction $\pi^{-}p\rightarrow \pi^{0}n$
in the momentum range from 103 to 178 MeV/$c$ }

\author{D. Mekterovi\'{c}}
\email[]{dmekter@irb.hr}
\affiliation{Ru\dd{}er Bo\v{s}kovi\'{c} Institute, Zagreb 10000, Croatia}
\author{I. Supek}
\affiliation{Ru\dd{}er Bo\v{s}kovi\'{c} Institute, Zagreb 10000, Croatia}
\author{V. Abaev}
\affiliation{Petersburg Nuclear Physics Institute, Gatchina 188350, Russia}
\author{V. Bekrenev}
\affiliation{Petersburg Nuclear Physics Institute, Gatchina 188350, Russia}
\author{C. Bircher}
\affiliation{Abilene Christian University, Abilene, Texas 79699-7963, USA}
\author{W. J. Briscoe}
\affiliation{The George Washington University, Washington, DC 20052-0001, USA}
\author{R. V. Cadman}
\thanks{Present address: Wisconsin Institutes for Medical Research, Madison, Wisconsin 53705, USA}
\affiliation{Argonne National Laboratory, Argonne, Illinois 60439-4815, USA}
\author{M. Clajus}
\affiliation{University of California, Los Angeles, California 90095-1547, USA}
\author{J. R. Comfort}
\affiliation{Arizona State University, Tempe, Arizona 85287-1504, USA}
\author{K. Craig}
\affiliation{Arizona State University, Tempe, Arizona 85287-1504, USA}
\author{D. Grosnick}
\affiliation{Valparaiso University, Valparaiso, Indiana 46383-6493, USA}
\author{D. Isenhover}
\affiliation{Abilene Christian University, Abilene, Texas 79699-7963, USA}
\author{M. Jerkins}
\affiliation{Abilene Christian University, Abilene, Texas 79699-7963, USA}
\author{M. Joy}
\affiliation{Abilene Christian University, Abilene, Texas 79699-7963, USA}
\author{N. Knecht}
\affiliation{University of Regina, Saskatchewan, Canada S4S OA2}
\author{D. D. Koetke}
\affiliation{Valparaiso University, Valparaiso, Indiana 46383-6493, USA}
\author{N. Kozlenko}
\affiliation{Petersburg Nuclear Physics Institute, Gatchina 188350, Russia}
\author{A. Kulbardis}
\affiliation{Petersburg Nuclear Physics Institute, Gatchina 188350, Russia}
\author{S. Kruglov}
\affiliation{Petersburg Nuclear Physics Institute, Gatchina 188350, Russia}
\author{G. Lolos}
\affiliation{University of Regina, Saskatchewan, Canada S4S OA2}
\author{I. Lopatin}
\affiliation{Petersburg Nuclear Physics Institute, Gatchina 188350, Russia}
\author{D. M. Manley}
\affiliation{Kent State University, Kent, Ohio 44242-0001, USA}
\author{R. Manweiler}
\affiliation{Valparaiso University, Valparaiso, Indiana 46383-6493, USA}
\author{A. Maru\v{s}i\'{c}}
\affiliation{University of California, Los Angeles, California 90095-1547, USA}
\author{S. McDonald}
\affiliation{University of California, Los Angeles, California 90095-1547, USA}
\author{B. M. K. Nefkens}
\affiliation{University of California, Los Angeles, California 90095-1547, USA}
\author{J. Olmsted}
\thanks{Present address: Lancaster General Hospital, Radiation Oncology, Lancaster, PA 17604 USA}
\affiliation{Kent State University, Kent, Ohio 44242-0001, USA}
\author{Z. Papandreou}
\affiliation{University of Regina, Saskatchewan, Canada S4S OA2}
\author{D. Peaslee}
\thanks{deceased}
\affiliation{University of Maryland, College Park, Maryland 20742-4111, USA}
\author{J. Peterson}
\affiliation{University of Colorado, Boulder, Colorado 80309-0390, USA}
\author{N. Phaisangittisakul}
\affiliation{University of California, Los Angeles, California 90095-1547, USA}
\author{S. N. Prakhov}
\affiliation{University of California, Los Angeles, California 90095-1547, USA}
\author{J. W. Price}
\affiliation{University of California, Los Angeles, California 90095-1547, USA}
\author{A. Ramirez}
\affiliation{Arizona State University, Tempe, Arizona 85287-1504, USA}
\author{M. E. Sadler}
\affiliation{Abilene Christian University, Abilene, Texas 79699-7963, USA}
\author{A. Shafi}
\affiliation{The George Washington University, Washington, DC 20052-0001, USA}
\author{H. Spinka}
\affiliation{Argonne National Laboratory, Argonne, Illinois 60439-4815, USA}
\author{S. Stanislaus}
\affiliation{Valparaiso University, Valparaiso, Indiana 46383-6493, USA}
\author{A. Starostin}
\affiliation{University of California, Los Angeles, California 90095-1547, USA}
\author{H. M. Staudenmaier}
\affiliation{Universit\"{a}t Karlsruhe, Karlsruhe 76128, Germany}
\author{I. Strakovsky}
\affiliation{The George Washington University, Washington, DC 20052-0001, USA}
\author{W. B. Tippens}
\affiliation{University of California, Los Angeles, California 90095-1547, USA}
\author{S. Watson}
\affiliation{Abilene Christian University, Abilene, Texas 79699-7963, USA}
\collaboration{Crystal Ball Collaboration}
\date{\today}

\begin{abstract}
Measured values of the differential cross sections for pion-nucleon charge 
exchange, $\pi^{-}p\rightarrow \pi^{0}n$, are presented for $\pi^{-}$ momenta 
of 103, 112, 120, 130, 139, 152, and 178~MeV/$c$. Complete angular distributions 
were obtained by using the Crystal Ball detector at the Alternating Gradient 
Synchrotron at Brookhaven National Laboratory. Statistical uncertainties 
of the differential cross sections vary from 3\% to 6\% in the backward 
angle region, and from 6\% to about 20\% in the forward region with the 
exception of the two most forward angles. The systematic uncertainties 
are estimated to be about 3\% for all momenta.
\end{abstract}
\maketitle
\section{Introduction \label{sec:Intro}}
There are three experimentally accessible scattering channels in the 
$\pi N$ system: $\pi^{+}p\rightarrow\pi^{+}p$ and 
$\pi^{-}p\rightarrow\pi^{-}p$ elastic scattering, and the 
$\pi^{-}p\rightarrow\pi^{0}n$ charge-exchange reaction (CEX). Precise data 
for all three channels are needed to obtain an accurate description of the 
$\pi N$ system via a consistent and complete set of scattering amplitudes. 
In that regard, the least satisfying situation is in the region below 
100~MeV $\pi^{-}$ kinetic energy where the experimental database for CEX 
is quite limited. On the other hand, the low-energy region is very 
interesting because isospin symmetry-breaking effects may be visible there.
 Furthermore, low-energy data have a strong impact on 
extrapolations of scattering amplitudes to the nonphysical region, and this 
extrapolation is needed to obtain important physical quantities such as 
the $\pi N$ $\sigma$ term.

Isospin symmetry is one of the key concepts in hadronic physics. Thus, it is 
very important to measure precisely small symmetry-breaking effects coming 
from both the up-down quark mass difference and the electromagnetic 
interaction. In the $\pi N$ system, one such isospin-breaking effect is a 
departure from the so-called triangle relation 
\begin{equation}
F_{CEX} = \frac{1}{\sqrt{2}}\left( F_{+} - F_{-} \right) \,,
\label{eq:tri}
\end{equation}
a relation between the scattering amplitudes for the CEX reaction and the two 
elastic channels. A surprisingly large 7\% violation of the triangle relation 
was obtained at 40~MeV in an analysis by Gibbs, Ai, and Kaufmann 
\cite{gibbs95}. As input, they used experimental data up to T$_{\pi}$ = 
50~MeV pion kinetic energy. A similar result was reported in independent work 
by Matsinos \cite{matsinos97,matsinos06} based on data up to T$_{\pi}$ = 
100~MeV. Fettes and Meissner \cite{fettes01} investigated isospin breaking in 
the framework of chiral perturbation theory up to 100~MeV/$c$ pion momentum and 
obtained only a 0.7\% effect. A similarly small effect was also found by 
the George Washington University (GWU) group, Gridnev {\it{et al.}} 
\cite{gridnev04}. Such a large discrepancy on this important question 
demands further investigation. From an experimental point of view, the 
weakest point is the lack of a sizeable set of data for low-energy CEX. 

Among the quantities extracted from low-energy $\pi N$ data, the $\pi N$ 
$\sigma$ term is distinguished both by its importance and by the lack of 
consensus on its precise value. For details, see \cite{schweitzer04} 
and references therein. 
The $\pi N$ $\sigma$ term is a measure of chiral symmetry breaking in the 
strong interaction. It is directly related to the strangeness content of 
the proton, and is also used in calculations of the mass spectra of hadrons 
and in searches for Higgs particles and dark matter. It is obtained by the 
extrapolation of the pion-nucleon scattering amplitudes to a 
negative energy point by taking advantage of their analytic properties. 
Charge-exchange data affect the determination of the $\sigma$ term 
indirectly, but are important to provide a stable database to determine 
the amplitudes as close to threshold as possible before extrapolating to 
the non-physical region.

The first CEX data below 100~MeV come from an experiment that used a single 
large NaI detector \cite{salomon84}, \cite{bagheri88}. The detector was 
placed at seven angles and collected continuous $\gamma$ energy spectra that 
were unfolded into CEX angular distributions, for six energies between 
27.4 and 91.7~MeV. Fitzgerald {\it{et al.}} \cite{fitzgerald86} used 
the Clinton P. Anderson Meson Physics Facility (LAMPF) $\pi^{0}$ spectrometer for 
coincident detection of the two $\pi^{0}$ decay photons. The experiment 
was performed at seven beam energies between 32.5 and 63.5~MeV and was restricted 
to laboratory polar angles smaller than 30$^{\circ}$. Later measurements 
were made with the same detector at 10, 20, and 40~MeV that covered a larger 
selection of forward and backward angles, but these results were reported 
only as incomplete and preliminary \cite{isenhower99}.  Frle\v{z} 
{\it{et al.}} obtained differential cross sections at 27.5~MeV in an 
angular range between 0$^{\circ}$ and 55$^{\circ}$ \cite{frlez98}. A 
previous experiment by the Crystal Ball Collaboration \cite{sadler04} 
measured differential cross sections in the region dominated by $\Delta$ 
resonance. The two lowest energies in that experiment overlap with the two 
highest energies reported in this work.

The most recent published results are from Jia {\it{et al.}} \cite{jia08}. 
They used the TRIUMF RMC spectrometer to obtain differential cross sections 
at six energies between 34.4 and 59.7~MeV in an angular range from 0$^{\circ}$ 
to 40$^{\circ}$. Their motivation was to check the results of Fitzgerald 
{\it{et al.}}, especially in the context of isospin violation. Serious 
disagreement between the two experiments was found. Finally, an experiment 
by Breitschopf {\it{et al.}} \cite{breitschopf06} should also be mentioned. 
They did not measure differential cross sections but did provide total CEX 
cross sections at many energies in the $\Delta$ region, nine of which are 
below 100~MeV.

The results from the new experiment described in this paper add 
differential cross sections with full angular coverage (20 different angles) 
for seven energies from 34 to 87~MeV.  Even if questions on the reliability 
of some of the previous experiments are ignored, these new data almost 
double the existing CEX data base below 100~MeV. Especially important is 
our contribution in the backward-angle region where we have small 
uncertainties and previous measurements are very scarce.
\section{Experimental Setup \label{sec:Exp_setup}}
The data presented in this work were measured in 2002 on the C6 beam line of 
the Alternating Gradient Synchrotron (AGS) at Brookhaven National Laboratory 
(BNL). The experiment (BNL experiment E958) was part of a program of baryon 
spectroscopy by the Crystal Ball Collaboration that included a long 
CEX run at higher energies in 1998. The experimental setups of 1998 and 
2002 were very similar (with the most important parts such as the Crystal 
Ball detector, the veto barrel, data acquisition, {\em etc.} being 
identical).  The experimental setup of the 1998 experiment is described in more detail 
in Ref. \cite{starostin01}.

\begin{figure}[p]
 \includegraphics[width=10cm]{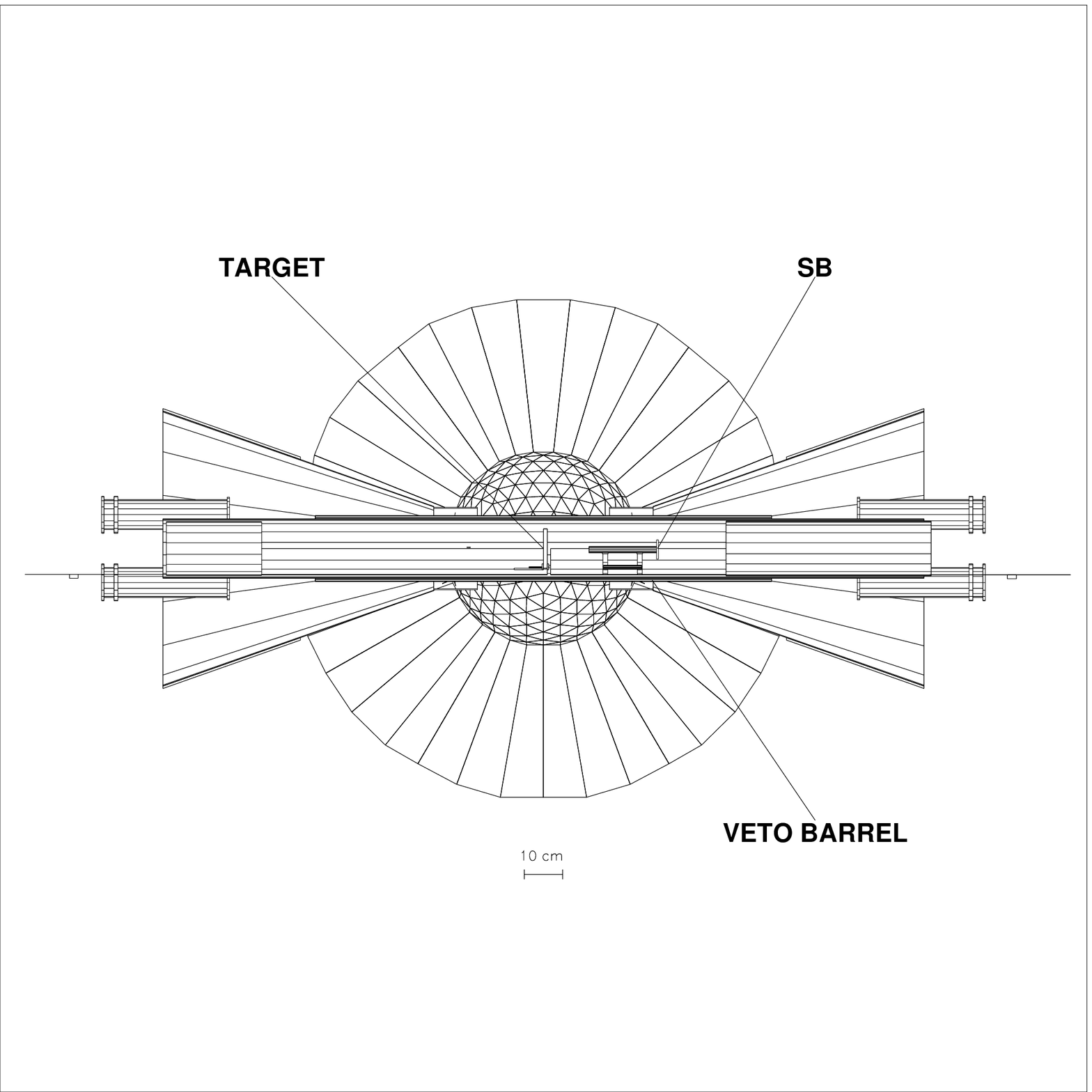}
 \caption{The section of the Crystal Ball detector along the beam axis and 
including other elements of the experimental setup such as the target, the 
SB counter, and the veto barrel.}
 \label{fig:CB}
 \end{figure}

The main part of the experimental setup is the Crystal Ball (CB) detector, 
illustrated in Fig.\ \ref{fig:CB}. 
It is an electromagnetic calorimeter and spectrometer in the shape of a ball 
with a cavity in the center (which housed a target), and openings for beam 
entrance and exit. The openings reduce the geometric acceptance to 93\% of 
4$\pi$ sr. The CB consists of 672 optically isolated NaI(Tl) crystals, each 
of them viewed by a single photomultiplier tube (PMT). The crystals are 
truncated triangular pyramids 5~cm on an edge at the inner radius, 13~cm at 
the outer radius, and 41~cm long.

The target used in the experiment was a cylinder 10~cm in diameter and 1~cm 
thick, composed of polyethylene (CH2). Its downstream face was positioned 
in the center of the CB cavity. Data were also taken with a carbon target 
of the same shape, size, and position in order to subtract the background 
produced by scattering and interaction of pions on carbon nuclei. Remaining
beam related backgrounds were subtracted by using data taken without a target.

A veto barrel (VB) was installed to reject events that had charged particles 
in the final state. It was constructed of four curved plastic scintillators 
that formed a cylindrical shell around the beam pipe. The beam pipe was a 
cylindrical aluminum support element placed through the beam opening of the 
CB. Each segment of the VB was 5~mm thick and 120~cm long. Each end of the four 
segments was viewed by a photomultiplier tube.

There were five plastic scintillators, a set of drift chambers, and a gas 
\v{C}erenkov counter along the beam line. The first scintillator S1 on the 
beam line was positioned before the last bending magnet D2. All signals 
were timed with respect to an ST scintillator that was placed 160~cm 
upstream from the target. Beam normalization was achieved with a small 
5-cm-by-5-cm scintillator, called SB, inserted 30~cm upstream from the 
target. All three scintillators were viewed by two PMTs (left and right). 
A coincidence between S1, ST, and SB defined the beam. For time-of-flight 
(TOF) analysis, another scintillator, designated BVS, was also used. 
It was positioned 210~cm downstream of the target. The \v{C}erenkov counter was
moved upstream from its position in the 1998 experiment and positioned just after 
the BVS. It was used to measure the electron contamination of the beam.
Beam trajectories were 
measured by the six drift chambers (three for the horizontal coordinate and 
three for the vertical coordinate). A drift chamber before D2 was used to 
determine the difference in momentum of the beam particle from the central 
value set by the beam tune.

Data collected with a pulser trigger (random trigger), a beam trigger, a 
charged trigger, and a neutral trigger were used in the analysis. The beam 
trigger was defined as a coincidence between signals from S1, ST, and SB:
\begin{equation}
Beam = S1\cdot ST\cdot SB \,.
\label{eq:beam_trig}
\end{equation}
Data collected with this trigger were used in beam normalization studies. 
Both the charged and the neutral trigger included the beam requirement 
mentioned above. In addition, they required that the total energy in the 
CB should be above 75~MeV. These triggers differed as to whether they included 
the existence of a signal from the veto barrel. Thus, the charged trigger 
was defined as
\begin{equation}
Ch. = Beam\cdot \left(E_{CB}>75\;\text{MeV} \right) \cdot VB \,,
\label{eq:ch_trig}
\end{equation} 
and the neutral trigger
\begin{equation}
Neut. = Beam\cdot \left(E_{CB}>75\;\text{MeV} \right) \cdot \overline{VB} \,.
\label{eq:neut_trig}
\end{equation} 
The charged trigger was needed for calibration of the VB, and the neutral 
trigger was our primary trigger from which cross sections were calculated. 
All triggers except the neutral trigger were prescaled.
\section{Data Analysis \label{sec:Analysis}}
The differential cross sections were calculated by using the expression
\begin{equation}
\frac{d\sigma}{d\Omega} = \frac{1}{\Delta\Omega t \epsilon}
\left[ \frac{Y_{CH2}}{N_{\pi^{-}}^{CH2}} - r_{c}r_{\epsilon}F\frac{Y_{C}}{N_{\pi^{-}}^{C}} - 
\left( 1 - r_{c}r_{\epsilon}F\right) \frac{Y_{E}}{N_{\pi^{-}}^{E}} \right] \,,
\label{eq:dcs}
\end{equation}
where the indices CH2, C, and E refer to data taken with the polyethylene 
target, carbon target, or no target, respectively; Y is the number of 
detected $\pi^{0}$s; $N_{\pi^{-}}$ is the number of beam $\pi^{-}$s
incident on a target and corrected for live time; $r_{c}$ is the ratio of 
the numbers of carbon nuclei in the CH2 and C targets; $r_{\epsilon}$ is 
the ratio of acceptances for scattering on carbon nuclei for the CH2 and 
C targets; $F$ is a small correction factor to compensate for differences 
between the average $\pi^{-}$ momenta in the CH2 and carbon targets;
$\epsilon$ is the acceptance for the CH2 target; $t$ is proton areal density 
in the CH2 target; and $\Delta\Omega$ is the size of the solid angle bin. 

The proton area density, $t$, and ratio $r_{c}$ were determined from the 
target specifications, and the size of the solid angle bin is given by 
the number of bins.

The CH2 and carbon targets had the same length, but the carbon target was 
more dense so the average $\pi^{-}$ momentum in it was lower than the 
average momentum for the CH2 target by 0.6~MeV/$c$ to 1.3~MeV/$c$, going from 
the highest to the lowest momentum. This shift led to a correction factor 
$F$: 
\begin{equation}
\frac{d\sigma_{C}}{d\Omega} \left(p_{CH2}\right) = 
  F \frac{d\sigma_{C}}{d\Omega} \left(p_{C}\right) \,,
\label{eq:F}
\end{equation}
which is a ratio of differential cross sections for CEX on carbon nuclei at 
the momenta in the CH2 and carbon targets.  The values for $F$ were estimated 
from the data for the carbon target. They were found to vary from about 1.005 
to about 1.05, going from the highest to the lowest momentum. The uncertainty 
of this estimate was taken to be $(F-1)/2$ and was included in the 
estimate of the systematic uncertainty. 
 
Calculation of the remaining quantities in Eq.\ \ref{eq:dcs}, together with 
the precise determination of beam momenta, is described in the following 
sections. As mentioned in the Introduction, our collaboration has already 
published CEX data, Sadler {\it{et al.}} \cite{sadler04}, in the $\Delta$ 
region, partially overlapping in energy with the experiment reported here. 
The analysis here is very similar to that of Sadler {\it{et al.}}, and we 
will refer to it for some aspects of the analysis.
\subsection{Detection of $\pi^{0}$s \label{sec:det_pi0}}
Data taken with the carbon target and no target were analyzed in the same 
way as data with the CH2 target and were subtracted from the CH2 data as 
given by Eq.\ \ref{eq:dcs}. In this way, the carbon background and accidental 
backgrounds were removed. The only other possible source of background 
would be pion radiative capture, $\pi^{-}p\rightarrow \gamma n$, but Monte Carlo 
analysis showed that it contributed less then 0.2\% and it was thus ignored.

The $\pi^{-}p\rightarrow \pi^{0}n$ reaction was identified in the subset of 
data collected with a neutral trigger by measuring the energy and direction 
of the two photons from $\pi^{0}\rightarrow \gamma\gamma$ decay. Each photon 
produced an electromagnetic shower in the CB that spread over several 
neighboring crystals. The first step in the analysis was to find such sets 
of crystals, called clusters. In a cluster-finding algorithm, a set of 
crystals with a deposited energy greater then 7~MeV was found. Such crystals, 
if not neighboring each other, were declared as a central crystal. A cluster 
was then defined to be the central crystal and its 21 neighbors with the 
condition that the total energy of the four highest-energy crystals in the 
cluster should be at least 17.5~MeV. The position of the cluster was the 
weighted average of positions of the crystals in the cluster, where the 
weighting factor was the square root of the deposited energy.

Once the energies and positions of all clusters were found, a number of cuts 
were applied to select the CEX events. Because detection of $\pi^{0}$s 
required detection of both photons from the decay, at least two clusters were 
needed. A third cluster could come from the neutron, so events with three 
clusters were included but events with more than three clusters were 
rejected. This cut suppressed accidental backgrounds. The most important 
cut, the signature of the CEX reaction, is that the invariant mass of only 
one pair of clusters (for 3-cluster events there are 3 pairs of clusters) be 
equal, within experimental resolution, to the $\pi^{0}$ mass. 

A very important cut is one on the missing mass because it strongly suppresses the 
carbon background. For CEX reactions on a proton target, the missing mass 
is equal to the neutron mass, while for the background reaction on carbon 
nuclei it is somewhat higher. Requiring that the missing mass be less than 
955~MeV/$c^{2}$ reduces the carbon background by about half and thus further 
reduces statistical and systematic uncertainties. This very strong cut 
changed the total cross section (due to a small difference between real and 
Monte Carlo data) by 1\% - 2\%. This change was included in the estimate 
of systematic uncertainties. Accidental backgrounds, because they were beam 
related, had the strongest impact on crystals in the region surrounding
the beam entrance or exit. In the final analysis, the beam entrance was 
excluded for clusters reconstructed to come from $\pi^{0}$. 

Other cuts designed to suppress backgrounds include one cut on the opening 
angle between clusters that form a candidate $\pi^{0}$, a cut on the energy 
of the CB excluding the tunnel region, and a cut that selects $\pi^{-}$ 
from TOF data. These cuts did not have strong impacts on the cross sections. 
The TOF cut was not only used to suppress backgrounds, but also to improve 
beam normalization as described in Sec.\ \ref{sec:bm_norm}. All tests used 
in reconstruction of $\pi^{0}$s are listed in Table \ref{tab:cuts}.
For events that passed all cuts, the center-of-mass (c.m.) scattering angle 
of the $\pi^{0}$ was calculated and histogramed into 20 bins of $\cos 
\theta_{c.m.}$.
 \begin{table*}
 \caption{Definitions of tests used in reconstruction of $\pi^{0}$s. 
Opening angle and TOF cuts are momentum dependent.}
 \label{tab:cuts}
 \begin{ruledtabular}
 \begin{tabular}{cc}
$\quad\quad$Test & Definition$\quad\quad$ \\
\hline
$\quad\quad$Neutral trigger & 
$S1\cdot ST\cdot SB\cdot \left(E_{CB}>75\;\text{MeV} \right) \cdot \overline{VB}$ $\quad\quad$ \\
$\quad\quad$Invariant mass &
 Only one cluster pair with: $105\;\text{MeV}/c^{2}<M_{I}<155\;\text{MeV}/c^{2}$ $\quad\quad$\\
$\quad\quad$Missing mass &
 For $\pi^{0}$ cluster pair: $917\;\text{MeV}/c^{2}<M_{M}<955\;\text{MeV}/c^{2}$ $\quad\quad$ \\
$\quad\quad$ Opening angle &
 For $\pi^{0}$ cluster pair: $\theta > \theta_{\text{min}}$ $\quad\quad$ \\
$\quad\quad$ Number of clusters &
 $N_{c}\le 3$ $\quad\quad$ \\
$\quad\quad$ Geometry cut  &
 Beam-in tunnel region excluded for $\pi^{0}$ clusters $\quad\quad$ \\
$\quad\quad$ CB energy without tunnel region &
$E_{CB}^{in}>80\;\text{MeV}$ $\quad\quad$ \\
$\quad\quad$ TOF &
$TDC_{SB}-TDC_{S1}>TDC_{\text{min}}$ $\quad\quad$
 \end{tabular}
 \end{ruledtabular}
 \end{table*}
\subsection{Monte Carlo and acceptance}
Two Monte Carlo simulation programs were used in the analysis, 
both of them based on GEANT 3 \cite{geant}.  The first program (BEAMLINE) 
simulated the passage of beam particles along the beam path. It was used 
for beam normalization studies as described in Section \ref{sec:bm_norm}. 
The BEAMLINE program provides a simple simulation with only a few elements 
along the beam path, and it tracks particles only inside a small tube 
along the beam axis. Our main simulation program CBALL includes a very 
detailed description of the experimental setup: all 672 crystals, the 
CB enclosure, the target assembly, the beam pipe, the veto barrel, and 
all scintillation counters in the trigger were included in the simulation. 

The CBALL simulation played a very important role in the analysis. In 
addition to the usual purpose of Monte Carlo simulations, to gain insight 
and confidence in the performance of the experimental setup, it was used to 
calculate the acceptance and the ratio of acceptances ($\epsilon$ and 
$r_{\epsilon}$ from Eq.\ \ref{eq:dcs}), to evaluate the fraction of events 
that would trigger the veto system (VB calibration), and to calibrate the 
beam momentum as discussed in Sec.\ \ref{sec:mom_calib}. Included in the 
input data for CBALL were distributions of CEX interaction points in the 
target and distributions of energies and directions of $\pi^{0}$s and 
neutrons. These input quantities were provided by a separate kinematics 
program DECKIN. This program selected interaction points randomly from 
among the measured beam trajectories saved from experimental data. The 
directions of $\pi^{0}$s were also selected randomly from a given angular 
distribution. The polar angular distribution was either flat (for final 
acceptance calculation) or given by a recent GWU partial-wave analysis 
(for momentum calibration). The remaining CBALL input, energies of 
$\pi^{0}$s and the energies and directions of neutrons, were calculated 
from two-body kinematics by using as input a distribution of $\pi^{-}$ 
momenta. The central value of the $\pi^{-}$ momentum was input, 
and the distribution around that central value was obtained from 
experimental data. Energy loss in the target was included in the above 
calculations.

\begin{figure*}[p]
 \includegraphics[width=18cm]{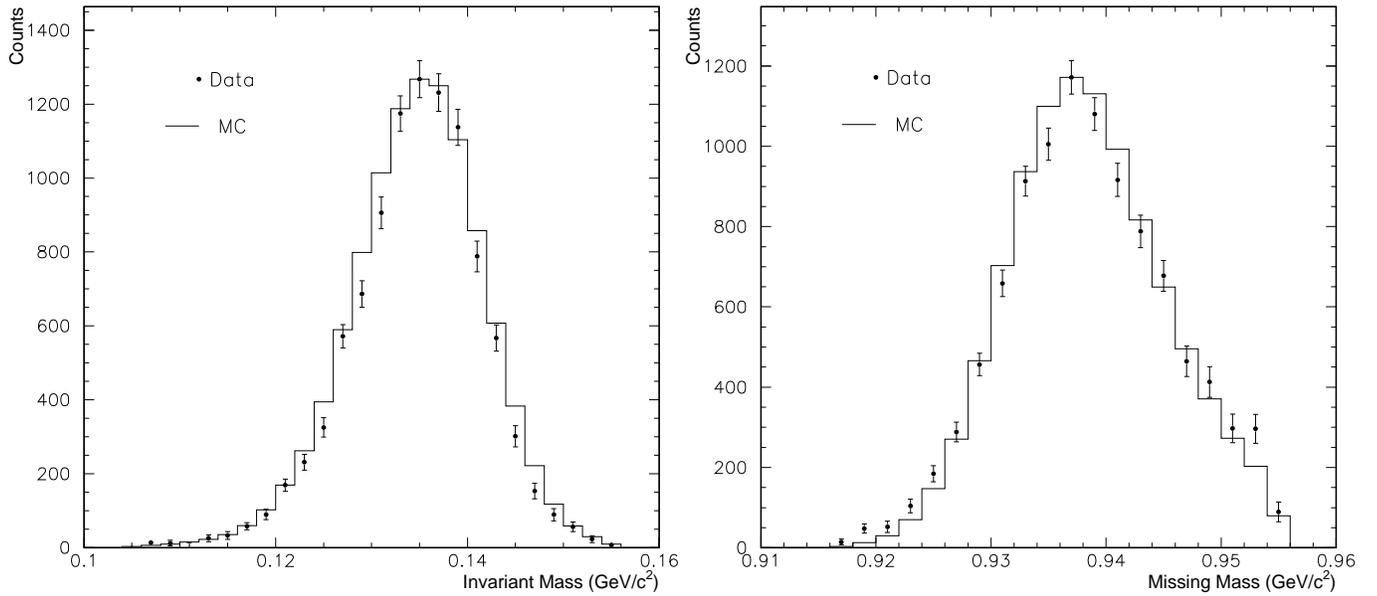}
 \caption{Comparison between real and Monte Carlo data for momentum 130~MeV/$c$
 of the invariant-mass and missing-mass distributions for pairs of 
clusters reconstructed to come from $\pi^{0}$ decays. The normalized empty 
and carbon target subtractions were applied to real data.}
 \label{fig:Inv-miss}
 \end{figure*}

\begin{figure}[p]
 \includegraphics[width=10cm]{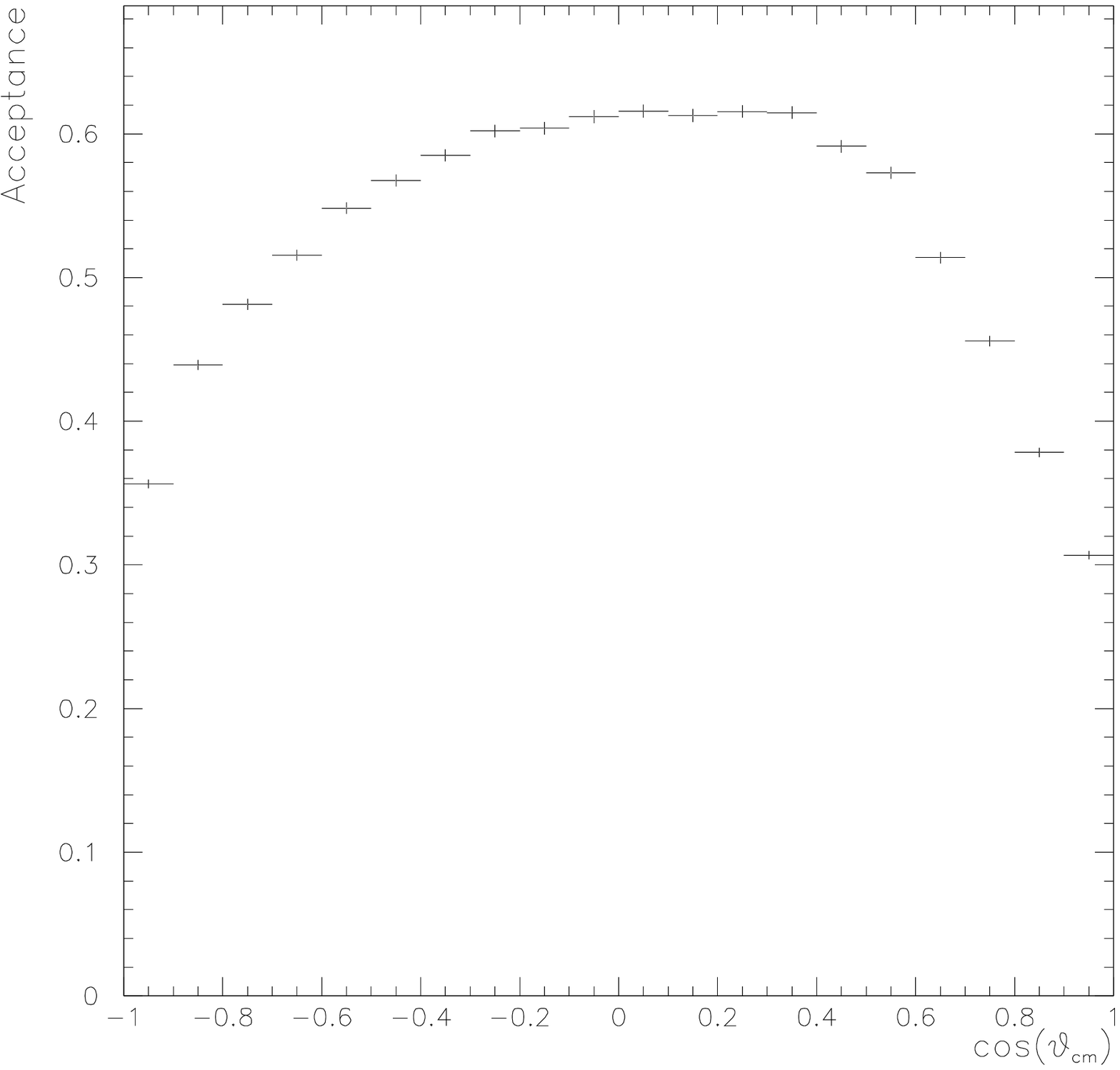}
 \caption{Acceptance for momentum 178~MeV/$c$.}
 \label{fig:acc}
 \end{figure}

The two photons from $\pi^{0}$ decay and the neutron were tracked by CBALL 
through all elements on which they were incident and the deposited energies 
were recorded. For crystals, energies recorded with the pulser trigger were 
randomly added to simulate accidental backgrounds. The Monte Carlo events 
generated in this way were then analyzed in the same way as the real data 
with the exception of the TOF cut, see Sec.\ \ref{sec:det_pi0}. Distributions 
of invariant mass and missing mass for real and Monte Carlo data of momentum 
130~MeV/$c$ are shown in Fig.\ \ref{fig:Inv-miss}.
The acceptance for a given bin was the ratio of the number of events that 
passed all cuts to the number generated. It is shown in Fig.\ \ref{fig:acc} 
for the momentum 178~MeV/$c$.
The ratio of acceptances for scattering on carbon nuclei in the CH2 and carbon 
targets, $r_{\epsilon}$, was also calculated from Monte Carlo data. 
As stated in connection with the correction factor $F$ in Sec.\ \ref{sec:Analysis},
average $\pi^{-}$ momentum in the carbon target was lower than the one in the CH2 target.
This difference in average momenta affected $r_{\epsilon}$ through the momentum dependence
of the missing mass cut, but even if the target
thicknesses were chosen such that average $\pi^{-}$ momentum was the same in both
targets, $r_{\epsilon}$ would still be needed because photons from the $\pi^{0}$ decay would still travel through
the materials of different density.
The values for $r_{\epsilon}$ ranged from about 0.94 to about 0.99
going from the lowest to the highest momentum.

The two photons could convert to $e^{+}e^{-}$ or the neutrons could 
interact hadronically before traversing the veto barrel. Such events were 
rejected by the veto barrel if the energy deposited exceeded the threshold. 
The veto barrel calibration and a discussion on reliability of Monte Carlo 
simulations is described in \cite{sadler04}.
\subsection{Beam normalization \label{sec:bm_norm}}
The number of beam particles $B$ was known at the position of SB (30~cm 
upstream from the target) because a coincidence between S1, ST, and SB was a 
trigger requirement. Not all of these beam particles were pions because the 
beam was contaminated with muons and electrons. To obtain the number of 
pions that hit the target ($N_{\pi^{-}}^{T}$, $T=CH2,C,E$ from 
Eq.\ \ref{eq:dcs}), one needs to find the fraction of pions in the beam at 
SB and the fraction of pions that survived from SB to the target. The fraction 
of pions in the beam at SB was determined with the TOF method because 
electrons, muons, and pions differ in time of flight between S1 and SB. 
The so-called decay muons cannot be distinguished from pions in this way. 
Decay muons originate from beam pion decays after the last beam channel 
magnet and thus fall under the pion S1-SB TOF peak. Contributions of decay 
muons to the pion TOF peak were determined from Monte Carlo simulations. 
Monte Carlo simulations were also used to obtain the fraction of pions that survived 
from SB to the target.

Taking all the above considerations into account, the number of pions that 
hit the target, corrected for computer live time, was evaluated as
\begin{equation}
N_{\pi^{-}} = B\cdot l\cdot f_{TOF}\cdot f_{\pi}\cdot s \,,
\label{eq:N_pi}
\end{equation}
where $B$ is the number of beam particles that hit SB (and S1 and ST), 
$l$ is computer live-time correction, $f_{TOF}$ is the fraction of particles 
in the pion peak of the S1-SB TOF spectra, $f_{\pi}$ is fraction of pions 
in the pion TOF peak, and $s$ is pion survival rate from SB to target. 
The computer live-time correction was calculated as the ratio of collected 
and triggered events, that is
\begin{equation}
l = \frac{\text{neutral events}}{\text{neutral triggers}} \,.
\label{eq:l}
\end{equation}

The TOF analysis was done by fitting the total S1-SB TOF spectrum with TOF 
spectra that included only electrons, muons, or pions. Samples of electrons, 
muons, and pions could be obtained by appropriate cuts on TOF spectra from 
other TDCs, BVS for example. In addition, signals from the \v{C}erenkov 
counter were used as a flag for electrons, and a signature for a CEX reaction 
was a flag for pions. Various combinations of electron, muon, and pion 
samples obtained in these ways were used to get a good understanding of 
systematic uncertainties.

\begin{figure}[p]
 \includegraphics[width=10cm]{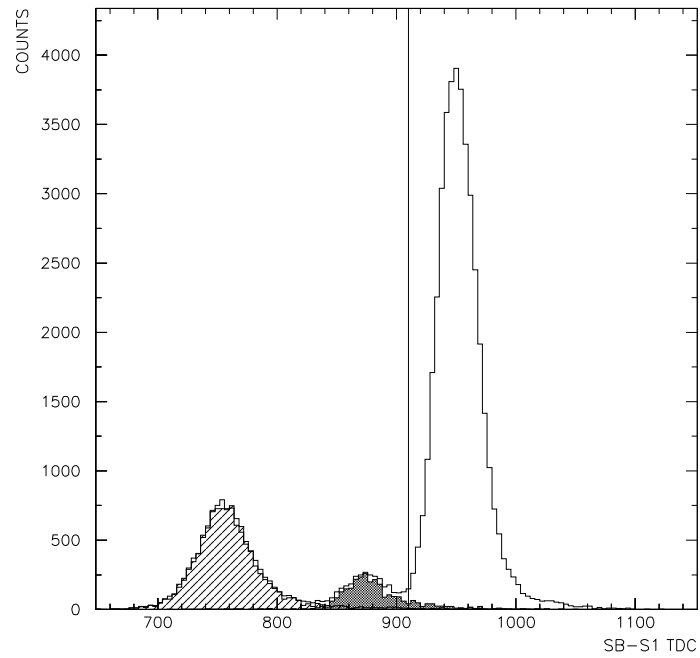}
 \caption{The difference between SB and S1 TDC values for data with $\pi^-$ 
momentum 120~MeV/$c$ and the CH2 target. The three peaks from left to right 
are from electrons, muons, and pions, respectively. Fits to the electron 
and muon samples are shown as shaded and black distributions, respectively. 
The position of the cut that selects the pion region is shown by a vertical 
line.}
 \label{fig:TOF}
 \end{figure}

The TOF spectra of the electron, muon, and pion samples were found to be 
asymmetric peaks with long tails towards increasing mass.  This fact 
means that electrons contributed to the muon peak and both electrons and 
muons contributed to the pion peak, but that the electron peak consisted 
almost entirely of electrons. Thus the procedure was defined as follows. 
The total TOF spectrum was fitted with the electron TOF spectrum in the 
region of electron dominance. The yield for the fitted electron spectrum 
divided by yield for the total spectrum represents the fraction of electrons 
in the beam. The fitted electron spectrum was then subtracted from the total 
spectrum. The remaining spectrum had only muon and pion peaks, and a region 
dominated by muons. That region was fitted with the muon TOF spectrum. 
As for electrons, the yield for the fitted muon spectrum divided by the yield 
for the total spectrum represents the fraction of muons in the beam. 
Having the electron and muon fractions gave the pion fraction. 
The fitted electron and muon TOF spectra for 120~MeV/$c$ are shown in Fig.\ \ref{fig:TOF} as 
shaded and black distributions, respectively, together with the total S1-SB 
TOF spectrum.

To reduce the systematic uncertainty of $f_{TOF}$, the pion fraction was 
determined only in the region dominated by pions. In Fig.\ \ref{fig:TOF}, 
this region is on the right side of the vertical line. Thus $f_{TOF}$
is a product of two factors. The first factor is the ratio of the number 
of events that had S1-SB TOF in the so-defined pion region to the total 
number of events. It was easily calculated and it only had a statistical 
uncertainty. The second factor is the fraction of pions in the defined 
pion region where the method described above was used to exclude decay 
muons.  Results for this latter number were very stable over different 
sets of data and for all except the highest momenta, varying between 
0.97 and 0.98. For the highest momenta, it was about 0.96. Calculating 
the pion fraction for only the pion region also means that the same cut 
had to be applied for the cross-section calculation. The cut 
eliminated only a small fraction of otherwise good candidates.  

As mentioned, the BEAMLINE simulation program was used to determine the 
contribution of decay muons and the fraction of pions that survived from SB 
to the target. Input for this program was a \emph{beam profile}, a value of 
the average (central) beam momentum, initial position of the beam, and the 
type of particle. By beam profile we refer to distributions of the positions 
(in a plane perpendicular to beam path), directions, and momenta (relative 
to the central momentum) of the beam particles. This information was 
obtained from the drift chambers for real data. Drift chambers were situated 
about 280~cm upstream from the target, so that position was chosen to be
the initial position of the beam simulation. From there, generated pions 
were tracked to the target.  Because the beam requirement was in the trigger, 
only events with pion or muon hits both in ST and SB were used in the 
analysis.  To check the influence of the beam profile on the final results, 
simulations were done separately by using beam profiles from individual 
runs. All differences found were within statistical uncertainties. Also, 
for every beam profile, two simulations that had different values of the 
average beam momentum were done. The average beam momenta at the initial 
position differed by 3~MeV/$c$ and were chosen to be around momenta obtained 
by momentum calibration. 

\begin{figure}[p]
 \includegraphics[width=10cm]{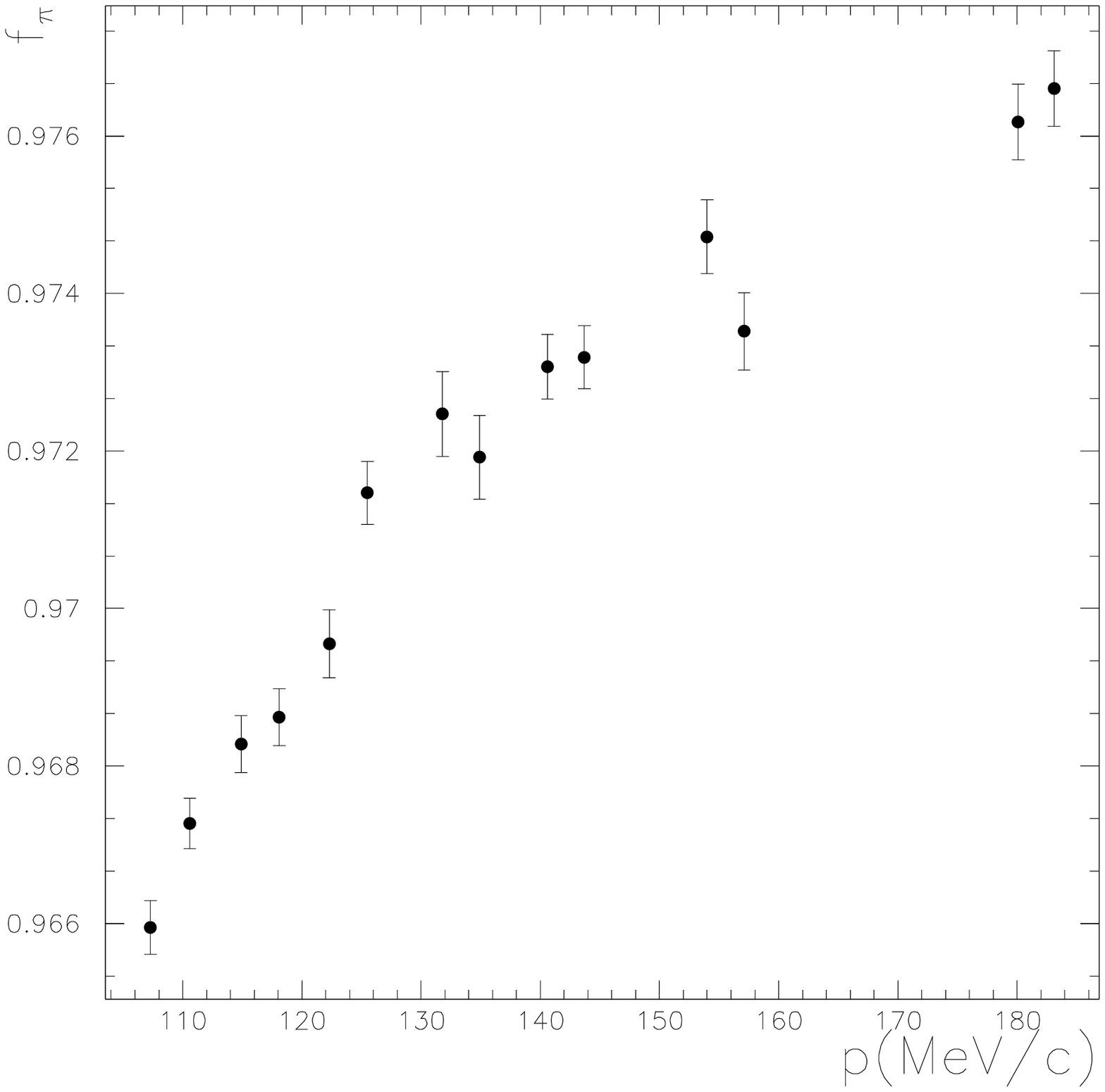}
 \caption{Fraction of pions in the set of particles with pion TOF at the 
position of SB, $f_{\pi}$ from Eq.\ \ref{eq:N_pi}, as a function of pion 
momentum at SB. Statistical uncertainties are shown with error bars.}
 \label{fig:f_pi}
 \end{figure}

\begin{figure}[p]
 \includegraphics[width=10cm]{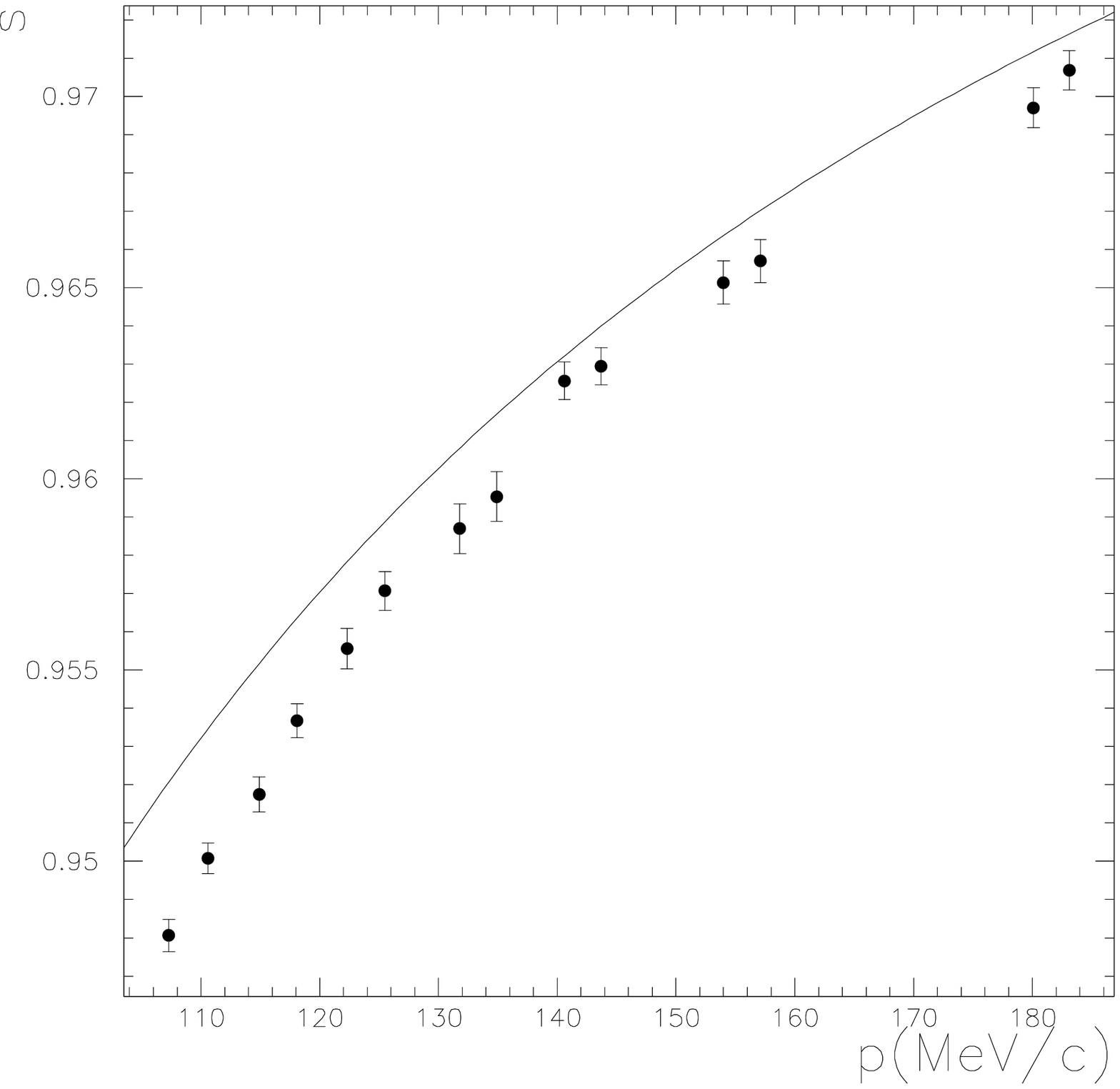}
 \caption{Survival rate of pions from SB to target, $s$ from Eq.\ 
\ref{eq:N_pi}, as a function of pion momentum at SB is shown with 
statistical uncertainties. The theoretical fraction of pions that do not 
decay in flight from SB to target is shown by a solid curve.}
 \label{fig:s}
 \end{figure}

Results for $f_{\pi}$ obtained by averaging simulations with beam profiles 
from all runs are shown in Fig.\ \ref{fig:f_pi} as a function of pion 
momentum at SB. 
In the same way as for $f_{\pi}$, results for $s$ are shown in Fig.\ \ref{fig:s}.
Also in Fig.\ \ref{fig:s} is shown with a solid curve the theoretical fraction 
of pions that do not decay from SB to target given by
\begin{equation} 
\label{eq:s_theory}
\frac{\pi^{th}_{T}}{\pi^{th}_{SB}} = \exp ( {-\frac{m}{c\tau}\frac{d}{p}} )\: \,,
\end{equation} 
where $c$ is the speed of light, $\tau$ is pion mean life, $m$ is pion mass, 
$d$ is the distance between SB and target, and $p$ is average pion momentum 
between SB and target. The difference between curve and survival rate 
represents scattered pions.
%
%
\subsection{Beam momentum calibration \label{sec:mom_calib}}
Beam momenta were obtained in three independent ways, which we call the 
previous calibration, calibration with invariant mass and missing mass 
(IM-MM calibration), and calibration with time-of-flight (TOF calibration).

The momentum calibration of the C6 beam line was checked extensively 
in previous experiments including ones by our collaboration 
\cite{starostin01,tippens01a,sadler04}. The beam momentum was thus 
approximately known at the position of the last dipole magnet, D2. A value 
of beam momentum at the target center can then be calculated by using this 
value as input for an energy-loss calculation. The main problem with this 
procedure is that previous beam-line calibrations were performed at higher 
momenta and then simply extrapolated to the momentum range of the current 
experiment, which can induce unknown systematic uncertainties.

The IM-MM calibration was based on comparing the invariant-mass and the 
missing-mass distributions for the real and Monte Carlo data. The average 
$\pi^{-}$ momentum at the target center was not known in the raw real data 
but could be assumed for the analysis. The missing-mass distribution depends 
on the assumed value of the $\pi^{-}$ momentum. For small relative changes, 
the dependence is linear. On the other hand, missing mass for the Monte 
Carlo data does not depend (or depends very weakly) on the beam momentum. 
The actual momentum is then equal to the assumed  momentum for which real 
and Monte Carlo missing-mass distributions are equal. For this procedure 
to be justified, real and Monte Carlo data have to be in good agreement. 
In particular, they have to have the same distribution in $\pi^{0}$ polar 
angle and they have to be gain-matched in CB energy. The procedure was as 
follows. Real distributions for invariant and missing mass were generated 
with the same cuts and background subtractions that were used for obtaining 
cross sections. Monte Carlo events, generated with the same cuts, were 
distributed in polar angle as predicted by the recent GWU analysis. 
Energy gain-matching was done so that real and Monte Carlo invariant mass 
peaks match. This match can be done either by adjusting the overall gain 
of the NaI crystals or the corresponding parameter in the Monte Carlo 
analysis. After real and Monte Carlo data were matched, the positions of the 
peaks in the missing-mass spectra were found with Gaussian fits for several 
different values of beam momentum. The uncertainty of this procedure was 
found to be 
\begin{equation} \label{eq:sigma_p}
\sigma_{P} \approx 5\sigma_{M}\: \,,
\end{equation} 
where $\sigma_{P}$ is the momentum uncertainty in MeV/$c$ and $\sigma_{M}$ 
is the uncertainty in the peak position of the mass distributions in 
MeV/$c^{2}$.  The uncertainty of the IM-MM calibration was estimated to be 
about 1.2~MeV/$c$.

In the TOF calibration, the theoretical and measured differences between 
pion and electron TOFs were compared. Electrons travel effectively at the 
speed of light so their TOF between two points on beam path is easily 
calculated. Pion TOF between the same two points was found for several 
different pion momenta, taking into account energy loss. In that way, 
a dependence between beam momentum and the difference in pion and electron 
TOFs was found. On the other hand, this difference was measured by finding 
(using Gaussian fitting) positions of pion and electron peaks in TOF spectra. 
The measured difference in TOF values suffers from two systematic uncertainties. 
The first source of uncertainty comes from an uncertainty of the precise value 
of the conversion factor between the time-to-digital converter (TDC) value 
and time. This uncertainty is not a limiting factor because a 1\% uncertainty in the 
conversion factor leads to $\approx$ 0.8~MeV/$c$ uncertainty in momentum. The second 
source of uncertainty is a known effect of electronic signal walk and it can be 
quite large. The uncertainty can be reduced by measuring TOF differences between 
several pairs of TDCs. The values of the S1 left and right TDCs were 
subtracted from: ST left and right, SB left and right, and the BVS TDC.  
The result is ten differences (six independent) in TDCs that were then used 
to form a weighted average where more weight was given to the differences in 
TDCs coming from the counters that were further apart. The exact procedure 
is not important because this method is less reliable than the IM-MM 
calibration method and it was used only as a consistency check. For the two 
highest momenta, the uncertainties of the procedure become too large to 
give useful results.

The average $\pi^{-}$ momenta at the target center were calculated in these 
three ways and are shown in the last three columns of Table~\ref{tab:mom_cal}. 
 \begin{table}
 \caption{Momentum calibration. The first two columns are the nominal and 
corrected momenta at D2. In third column are momenta at the target center 
obtained with a previous D2 magnet calibration and energy loss calculations, 
with input from the second column. The last two columns are momenta at the 
target center from the IM-MM calibration, and calibration from the TOF 
method. The TOF calibration can not be used for the two highest momenta.}
 \label{tab:mom_cal}
 \begin{ruledtabular}
 \begin{tabular}{ccccc}
$P_{D2}$ & $P_{D2}$ corr. & Prev. $P_{T}$ & IM-MM $P_{T}$ & TOF $P_{T}$\\
 \hline
119 & 117.5 & 107.2 & 103.0 & 105.2 \\
126 & 126.0 & 116.9 & 112.4 & 113.3 \\
132 & 131.3 & 122.8 & 120.0 & 121.4 \\
140 & 140.2 & 132.5 & 129.6 & 131.4 \\
148 & 148.5 & 141.4 & 139.2 & 139.6 \\
161 & 161.0 & 154.6 & 152.0 & -     \\
186 & 186.4 & 179.9 & 178.1 & -
 \end{tabular}
 \end{ruledtabular}
 \end{table}
Momenta from the previous calibration are systematically higher than momenta 
obtained with the TOF analysis, and the TOF momenta are systematically higher 
then momenta from the IM-MM calibration. The difference between the previous 
calibration and IM-MM tends to increase with decreasing momentum.  The IM-MM 
calibration was accepted as the most reliable. For the final results, IM-MM 
values (rounded to the nearest integer) and uncertainty estimate were used. 
\section{Results \label{sec:Res}}
\subsection{Systematic uncertainties \label{sec:sys_err}}
The following sources of systematic uncertainties were identified: beam 
normalization, calibration of the veto barrel, difference between the 
densities of the CH2 and carbon targets, and small differences between real 
and Monte Carlo data (cut sensitivity).

The uncertainty in beam normalization, as seen from Eq.\ \ref{eq:N_pi}, 
comes from uncertainties of beam Monte Carlo simulations (0.4\% - 0.2\%), 
statistical uncertainties of computer live time (0.3\% - 0.1\%), and 
statistical (0.4\% - 0.1\%) and systematic uncertainties (0.3\% - 1.0\%) in 
the determination of the pion fraction, $f_{TOF}$. The overall contribution 
of beam normalization to the total uncertainty was estimated to be 
0.7\% - 1.8\%. In the quoted ranges of uncertainties, the first number 
refers to the lowest momentum and second to the highest. 

The uncertainty of the veto barrel calibration was estimated to be 2\% for 
all momenta.

The difference between densities of the CH2 and carbon targets adds to the 
total uncertainty through the uncertainty in estimating the correction factor 
$F$ and the uncertainty of $r_{\epsilon}$. These two factors combined gave 
a 0.8\% - 1.3\% uncertainty.

Finally, the change in the total cross section with and without the 
missing-mass cut was a measure of systematic uncertainty coming from small 
differences between real and Monte Carlo data. For all momenta except the 
lowest, the uncertainty was about 1\%, and for the lowest it was about 2.5\%.

The total systematical uncertainty, obtained by adding in quadrature all of 
the described contributions, was close to 3\% for all momenta. The results 
are summarized in Table \ref{tab:tot_sigma}. 
 \begin{table*}[tbh!]
 \caption{Sources of systematical uncertainties and total systematical 
uncertainty.}
 \label{tab:tot_sigma}
 \begin{ruledtabular}
 \begin{tabular}{ccccccc}
Momentum & Kin.\ Energy & Beam          & Veto barrel & Difference in & Cut          & Total       \\
 (MeV/$c$) & (MeV)        & normalization & calibration &target densities       & sensitivity  & uncertainty \\
\hline
103      & 33.9         & 0.7\%         & 2.0\%       & 0.8\%         & 2.5\%        & 3.4\%       \\
112      & 39.4         & 0.9\%         & 2.0\%       & 1.0\%         & 1.0\%        & 2.6\%       \\
120      & 44.5         & 1.0\%         & 2.0\%       & 1.1\%         & 1.0\%        & 2.7\%       \\
130      & 51.2         & 1.3\%         & 2.0\%       & 1.2\%         & 1.0\%        & 2.9\%       \\
139      & 57.4         & 1.4\%         & 2.0\%       & 1.3\%         & 1.0\%        & 2.9\%       \\
152      & 66.8         & 1.7\%         & 2.0\%       & 1.1\%         & 0.8\%        & 3.0\%       \\
178      & 86.6         & 1.8\%         & 2.0\%       & 0.8\%         & 0.8\%        & 2.9\%
 \end{tabular}
 \end{ruledtabular}
 \end{table*}
\subsection{Cross sections \label{sec:cross_sec}}
The obtained values of $\pi^{-}p\rightarrow\pi^{0}n$ differential cross 
sections and their statistical uncertainties are listed in Tables~\ref{tab:res1} and \ref{tab:res2}.
 \begin{table*}[p]
 \caption{Differential cross sections.}
 \label{tab:res1}
 \begin{ruledtabular}
 \begin{tabular}{lcccc}
Momentum & 103~MeV/$c$ & 112~MeV/$c$ & 120~MeV/$c$ & 130~MeV/$c$\\
$\cos \theta_{c.m.}$ & $d\sigma/d\Omega$ $\quad$ uncertainty& $d\sigma/d\Omega$ $\quad$ uncertainty& $d\sigma/d\Omega$ $\quad$ uncertainty& $d\sigma/d\Omega$ $\quad$ uncertainty\\
\hline
-0.95 & 1.016 $\quad\quad\;\;\:$ 0.064$\;\;\;\;$& 1.145 $\quad\quad\;\;\:$ 0.062$\;\;\;\;$& 1.206 $\quad\quad\;\;\:$ 0.062$\;\;\;\;$& 1.567 $\quad\quad\;\;\:$ 0.074$\;\;\;\;$\\
-0.85 & 0.987 $\quad\quad\;\;\:$ 0.056$\;\;\;\;$& 1.045 $\quad\quad\;\;\:$ 0.055$\;\;\;\;$& 1.294 $\quad\quad\;\;\:$ 0.054$\;\;\;\;$& 1.469 $\quad\quad\;\;\:$ 0.066$\;\;\;\;$\\
-0.75 & 0.871 $\quad\quad\;\;\:$ 0.051$\;\;\;\;$& 1.028 $\quad\quad\;\;\:$ 0.050$\;\;\;\;$& 1.162 $\quad\quad\;\;\:$ 0.049$\;\;\;\;$& 1.299 $\quad\quad\;\;\:$ 0.059$\;\;\;\;$\\
-0.65 & 0.820 $\quad\quad\;\;\:$ 0.047$\;\;\;\;$& 0.892 $\quad\quad\;\;\:$ 0.044$\;\;\;\;$& 0.979 $\quad\quad\;\;\:$ 0.045$\;\;\;\;$& 1.129 $\quad\quad\;\;\:$ 0.053$\;\;\;\;$\\
-0.55 & 0.754 $\quad\quad\;\;\:$ 0.044$\;\;\;\;$& 0.793 $\quad\quad\;\;\:$ 0.042$\;\;\;\;$& 0.797 $\quad\quad\;\;\:$ 0.039$\;\;\;\;$& 1.021 $\quad\quad\;\;\:$ 0.048$\;\;\;\;$\\
-0.45 & 0.714 $\quad\quad\;\;\:$ 0.041$\;\;\;\;$& 0.798 $\quad\quad\;\;\:$ 0.040$\;\;\;\;$& 0.776 $\quad\quad\;\;\:$ 0.037$\;\;\;\;$& 0.857 $\quad\quad\;\;\:$ 0.043$\;\;\;\;$\\
-0.35 & 0.618 $\quad\quad\;\;\:$ 0.040$\;\;\;\;$& 0.642 $\quad\quad\;\;\:$ 0.036$\;\;\;\;$& 0.751 $\quad\quad\;\;\:$ 0.034$\;\;\;\;$& 0.799 $\quad\quad\;\;\:$ 0.040$\;\;\;\;$\\
-0.25 & 0.539 $\quad\quad\;\;\:$ 0.034$\;\;\;\;$& 0.593 $\quad\quad\;\;\:$ 0.033$\;\;\;\;$& 0.588 $\quad\quad\;\;\:$ 0.031$\;\;\;\;$& 0.640 $\quad\quad\;\;\:$ 0.037$\;\;\;\;$\\
-0.15 & 0.421 $\quad\quad\;\;\:$ 0.031$\;\;\;\;$& 0.493 $\quad\quad\;\;\:$ 0.031$\;\;\;\;$& 0.509 $\quad\quad\;\;\:$ 0.028$\;\;\;\;$& 0.563 $\quad\quad\;\;\:$ 0.032$\;\;\;\;$\\
-0.05 & 0.421 $\quad\quad\;\;\:$ 0.030$\;\;\;\;$& 0.429 $\quad\quad\;\;\:$ 0.028$\;\;\;\;$& 0.459 $\quad\quad\;\;\:$ 0.027$\;\;\;\;$& 0.545 $\quad\quad\;\;\:$ 0.030$\;\;\;\;$\\
0.05 & 0.415 $\quad\quad\;\;\:$ 0.028$\;\;\;\;$& 0.387 $\quad\quad\;\;\:$ 0.025$\;\;\;\;$& 0.417 $\quad\quad\;\;\:$ 0.024$\;\;\;\;$& 0.430 $\quad\quad\;\;\:$ 0.028$\;\;\;\;$\\
0.15 & 0.275 $\quad\quad\;\;\:$ 0.025$\;\;\;\;$& 0.318 $\quad\quad\;\;\:$ 0.024$\;\;\;\;$& 0.325 $\quad\quad\;\;\:$ 0.022$\;\;\;\;$& 0.368 $\quad\quad\;\;\:$ 0.026$\;\;\;\;$\\
0.25 & 0.245 $\quad\quad\;\;\:$ 0.023$\;\;\;\;$& 0.232 $\quad\quad\;\;\:$ 0.023$\;\;\;\;$& 0.239 $\quad\quad\;\;\:$ 0.021$\;\;\;\;$& 0.286 $\quad\quad\;\;\:$ 0.022$\;\;\;\;$\\
0.35 & 0.223 $\quad\quad\;\;\:$ 0.022$\;\;\;\;$& 0.201 $\quad\quad\;\;\:$ 0.021$\;\;\;\;$& 0.202 $\quad\quad\;\;\:$ 0.019$\;\;\;\;$& 0.227 $\quad\quad\;\;\:$ 0.020$\;\;\;\;$\\
0.45 & 0.188 $\quad\quad\;\;\:$ 0.021$\;\;\;\;$& 0.149 $\quad\quad\;\;\:$ 0.019$\;\;\;\;$& 0.154 $\quad\quad\;\;\:$ 0.016$\;\;\;\;$& 0.140 $\quad\quad\;\;\:$ 0.018$\;\;\;\;$\\
0.55 & 0.132 $\quad\quad\;\;\:$ 0.020$\;\;\;\;$& 0.138 $\quad\quad\;\;\:$ 0.017$\;\;\;\;$& 0.081 $\quad\quad\;\;\:$ 0.015$\;\;\;\;$& 0.130 $\quad\quad\;\;\:$ 0.018$\;\;\;\;$\\
0.65 & 0.125 $\quad\quad\;\;\:$ 0.019$\;\;\;\;$& 0.064 $\quad\quad\;\;\:$ 0.015$\;\;\;\;$& 0.111 $\quad\quad\;\;\:$ 0.015$\;\;\;\;$& 0.086 $\quad\quad\;\;\:$ 0.014$\;\;\;\;$\\
0.75 & 0.097 $\quad\quad\;\;\:$ 0.015$\;\;\;\;$& 0.056 $\quad\quad\;\;\:$ 0.014$\;\;\;\;$& 0.048 $\quad\quad\;\;\:$ 0.012$\;\;\;\;$& 0.070 $\quad\quad\;\;\:$ 0.014$\;\;\;\;$\\
0.85 & 0.062 $\quad\quad\;\;\:$ 0.014$\;\;\;\;$& 0.045 $\quad\quad\;\;\:$ 0.013$\;\;\;\;$& 0.043 $\quad\quad\;\;\:$ 0.011$\;\;\;\;$& 0.030 $\quad\quad\;\;\:$ 0.012$\;\;\;\;$\\
0.95 & 0.015 $\quad\quad\;\;\:$ 0.011$\;\;\;\;$& 0.014 $\quad\quad\;\;\:$ 0.011$\;\;\;\;$& 0.008 $\quad\quad\;\;\:$ 0.010$\;\;\;\;$& 0.020 $\quad\quad\;\;\:$ 0.012$\;\;\;\;$
 \end{tabular}
 \end{ruledtabular}
 \end{table*}
 \begin{figure*}[p!]
 \includegraphics[width=18cm]{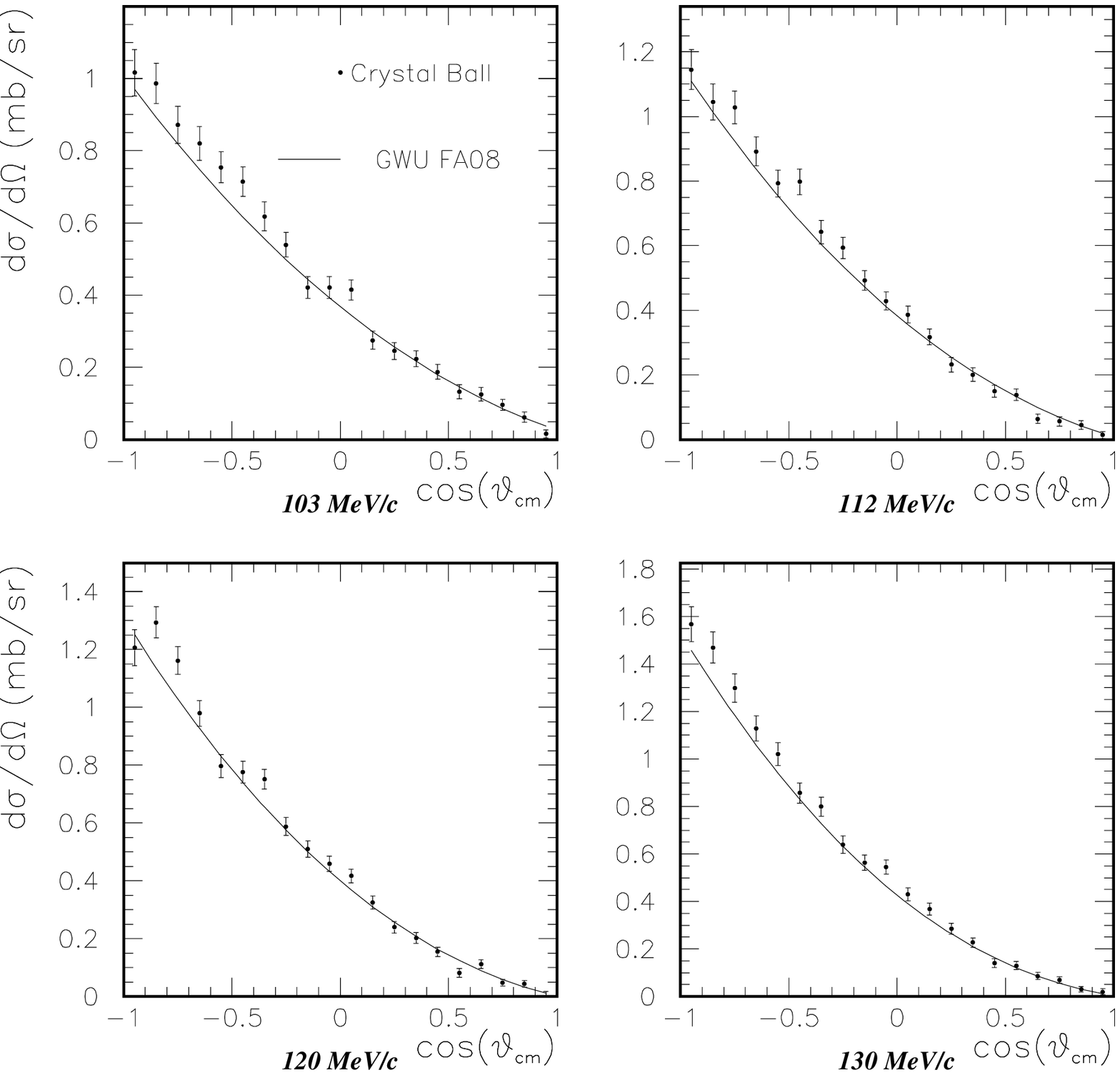}
 \caption{Differential cross sections of reaction 
$\pi^{-}p\rightarrow\pi^{0}n$. Black circles are the values obtained in this 
experiment with the error bars showing statistical uncertainties. The curves 
are the results of the FA08 partial-wave analysis of the George Washington 
group \cite{arndt06}.}
 \label{fig:dcs_4}
 \end{figure*}
 \begin{table*}[p!]
 \caption{Differential cross sections.}
 \label{tab:res2}
 \begin{ruledtabular}
 \begin{tabular}{lccc}
Momentum & 139~MeV/$c$ & 152~MeV/$c$ & 178~MeV/$c$\\
$\cos \theta_{c.m.}$ & $d\sigma/d\Omega$ $\quad$ uncertainty& $d\sigma/d\Omega$ $\quad$ uncertainty& $d\sigma/d\Omega$ $\quad$ uncertainty\\
\hline
-0.95 & 1.938 $\quad\quad\;\;\:$ 0.063$\;\;\;\;$& 2.157 $\quad\quad\;\;\:$ 0.073$\;\;\;\;$& 3.010 $\quad\quad\;\;\:$ 0.081$\;\;\;\;$\\
-0.85 & 1.587 $\quad\quad\;\;\:$ 0.055$\;\;\;\;$& 1.872 $\quad\quad\;\;\:$ 0.065$\;\;\;\;$& 2.609 $\quad\quad\;\;\:$ 0.068$\;\;\;\;$\\
-0.75 & 1.564 $\quad\quad\;\;\:$ 0.050$\;\;\;\;$& 1.768 $\quad\quad\;\;\:$ 0.057$\;\;\;\;$& 2.454 $\quad\quad\;\;\:$ 0.062$\;\;\;\;$\\
-0.65 & 1.306 $\quad\quad\;\;\:$ 0.045$\;\;\;\;$& 1.533 $\quad\quad\;\;\:$ 0.052$\;\;\;\;$& 2.163 $\quad\quad\;\;\:$ 0.056$\;\;\;\;$\\
-0.55 & 1.155 $\quad\quad\;\;\:$ 0.041$\;\;\;\;$& 1.353 $\quad\quad\;\;\:$ 0.046$\;\;\;\;$& 1.875 $\quad\quad\;\;\:$ 0.050$\;\;\;\;$\\
-0.45 & 1.042 $\quad\quad\;\;\:$ 0.036$\;\;\;\;$& 1.223 $\quad\quad\;\;\:$ 0.042$\;\;\;\;$& 1.589 $\quad\quad\;\;\:$ 0.045$\;\;\;\;$\\
-0.35 & 0.878 $\quad\quad\;\;\:$ 0.032$\;\;\;\;$& 1.021 $\quad\quad\;\;\:$ 0.039$\;\;\;\;$& 1.311 $\quad\quad\;\;\:$ 0.041$\;\;\;\;$\\
-0.25 & 0.719 $\quad\quad\;\;\:$ 0.029$\;\;\;\;$& 0.819 $\quad\quad\;\;\:$ 0.035$\;\;\;\;$& 1.127 $\quad\quad\;\;\:$ 0.036$\;\;\;\;$\\
-0.15 & 0.622 $\quad\quad\;\;\:$ 0.027$\;\;\;\;$& 0.717 $\quad\quad\;\;\:$ 0.031$\;\;\;\;$& 0.899 $\quad\quad\;\;\:$ 0.033$\;\;\;\;$\\
-0.05 & 0.502 $\quad\quad\;\;\:$ 0.024$\;\;\;\;$& 0.596 $\quad\quad\;\;\:$ 0.028$\;\;\;\;$& 0.771 $\quad\quad\;\;\:$ 0.030$\;\;\;\;$\\
 0.05 & 0.423 $\quad\quad\;\;\:$ 0.022$\;\;\;\;$& 0.491 $\quad\quad\;\;\:$ 0.025$\;\;\;\;$& 0.650 $\quad\quad\;\;\:$ 0.027$\;\;\;\;$\\
 0.15 & 0.334 $\quad\quad\;\;\:$ 0.019$\;\;\;\;$& 0.398 $\quad\quad\;\;\:$ 0.023$\;\;\;\;$& 0.506 $\quad\quad\;\;\:$ 0.024$\;\;\;\;$\\
 0.25 & 0.261 $\quad\quad\;\;\:$ 0.017$\;\;\;\;$& 0.280 $\quad\quad\;\;\:$ 0.020$\;\;\;\;$& 0.450 $\quad\quad\;\;\:$ 0.021$\;\;\;\;$\\
 0.35 & 0.221 $\quad\quad\;\;\:$ 0.016$\;\;\;\;$& 0.216 $\quad\quad\;\;\:$ 0.018$\;\;\;\;$& 0.374 $\quad\quad\;\;\:$ 0.020$\;\;\;\;$\\
 0.45 & 0.189 $\quad\quad\;\;\:$ 0.015$\;\;\;\;$& 0.174 $\quad\quad\;\;\:$ 0.017$\;\;\;\;$& 0.383 $\quad\quad\;\;\:$ 0.019$\;\;\;\;$\\
 0.55 & 0.114 $\quad\quad\;\;\:$ 0.013$\;\;\;\;$& 0.152 $\quad\quad\;\;\:$ 0.015$\;\;\;\;$& 0.322 $\quad\quad\;\;\:$ 0.017$\;\;\;\;$\\
 0.65 & 0.094 $\quad\quad\;\;\:$ 0.012$\;\;\;\;$& 0.142 $\quad\quad\;\;\:$ 0.015$\;\;\;\;$& 0.361 $\quad\quad\;\;\:$ 0.019$\;\;\;\;$\\
 0.75 & 0.072 $\quad\quad\;\;\:$ 0.012$\;\;\;\;$& 0.088 $\quad\quad\;\;\:$ 0.014$\;\;\;\;$& 0.329 $\quad\quad\;\;\:$ 0.019$\;\;\;\;$\\
 0.85 & 0.046 $\quad\quad\;\;\:$ 0.011$\;\;\;\;$& 0.114 $\quad\quad\;\;\:$ 0.014$\;\;\;\;$& 0.339 $\quad\quad\;\;\:$ 0.020$\;\;\;\;$\\
 0.95 & 0.038 $\quad\quad\;\;\:$ 0.011$\;\;\;\;$& 0.100 $\quad\quad\;\;\:$ 0.015$\;\;\;\;$& 0.369 $\quad\quad\;\;\:$ 0.024$\;\;\;\;$
 \end{tabular}
 \end{ruledtabular} 
 \end{table*} 
 \begin{figure*}[p]
 \includegraphics[width=18cm]{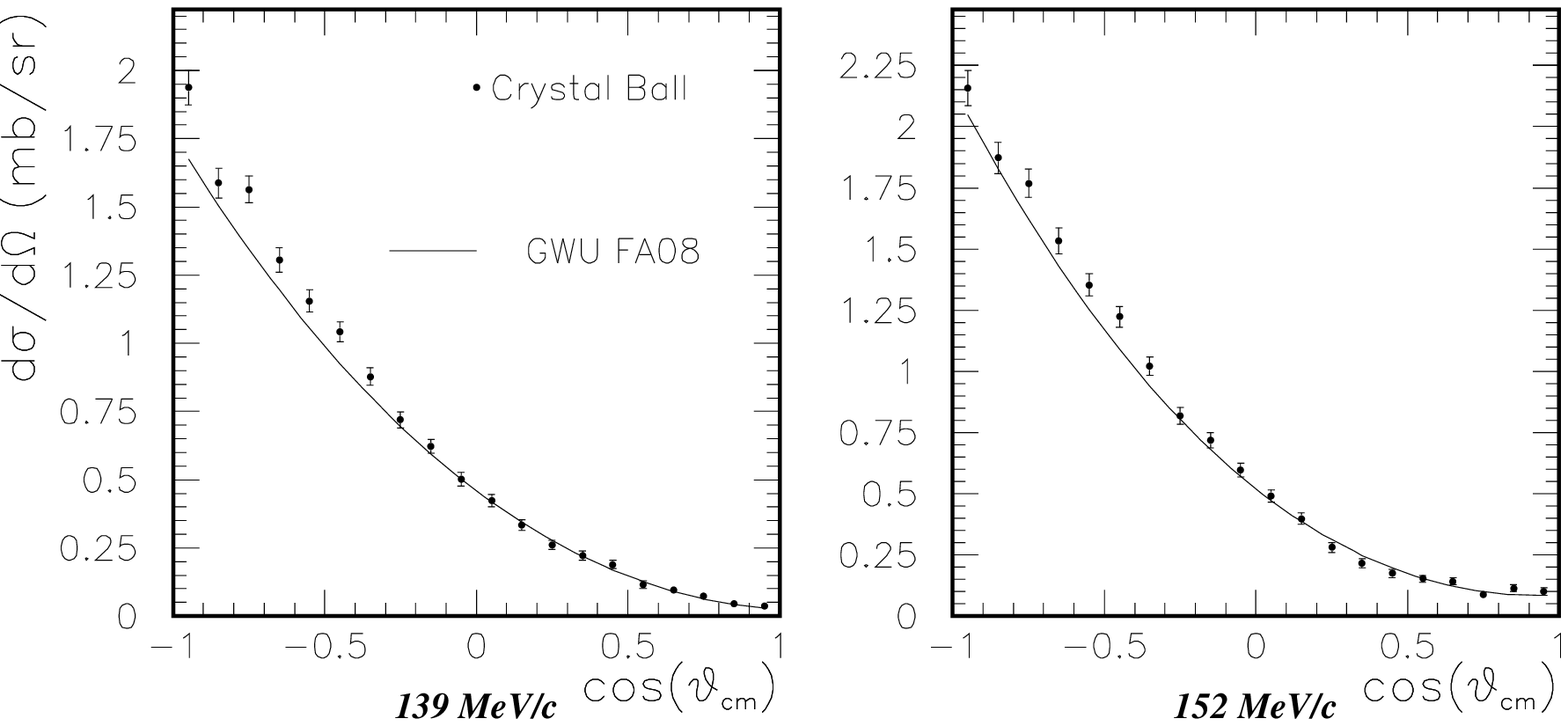}
 \caption{Differential cross sections for the reaction 
$\pi^{-}p\rightarrow\pi^{0}n$. Black circles are the values obtained in 
this experiment, with the error bars showing statistical uncertainties. 
The curves are  the results of the FA08 partial-wave analysis of the 
George Washington group \cite{arndt06}.}
 \label{fig:dcs_2}
 \end{figure*}
 \begin{figure}[p]
 \includegraphics[width=10cm]{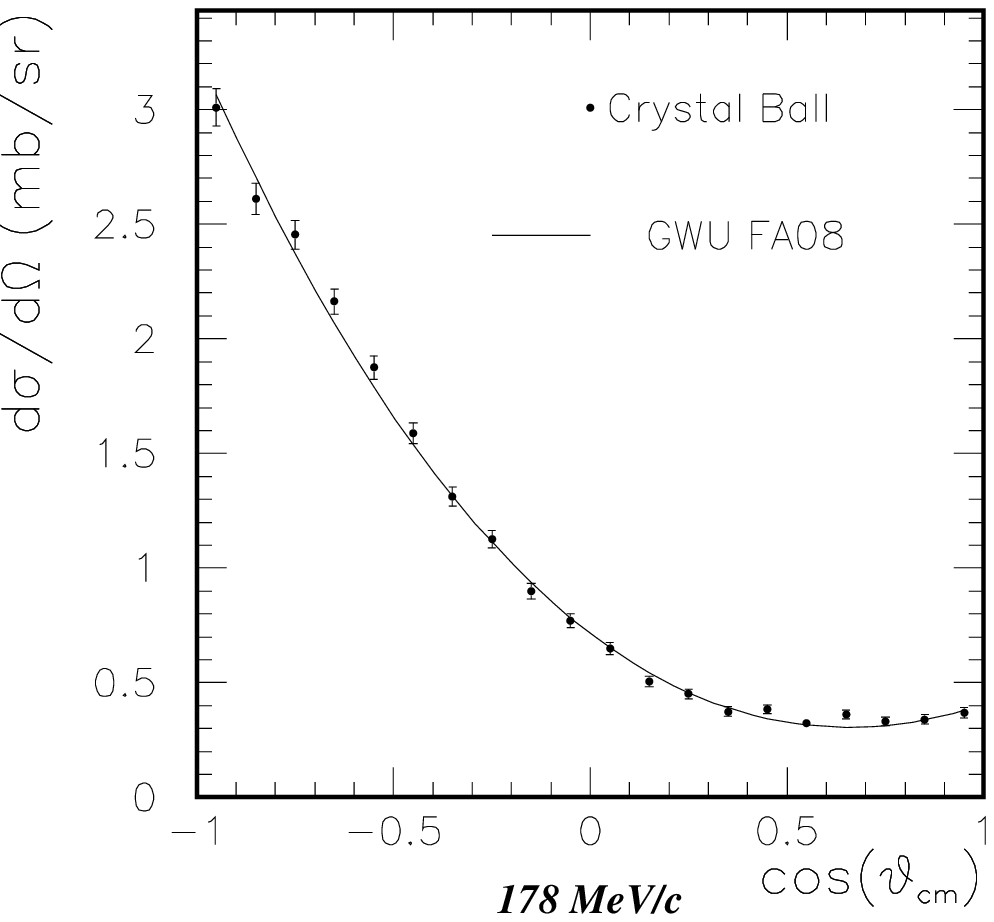}
 \caption{Differential cross sections of the reaction 
$\pi^{-}p\rightarrow\pi^{0}n$. Black circles are the values obtained in 
this experiment with the error bars showing statistical uncertainties. 
The curves are  the results of the FA08 partial-wave analysis of the 
George Washington group \cite{arndt06}.}
 \label{fig:dcs_1}
 \end{figure}

They are also plotted in Figs.\ \ref{fig:dcs_4}, \ref{fig:dcs_2}, and 
\ref{fig:dcs_1} together with the results of the FA08 partial-wave analysis 
(PWA) of the George Washington group \cite{arndt06}. Statistical 
uncertainties range from 3\% to 6\% for the highest momentum, to 6\% to 
15\% for the lowest momenta except for the few most forward-angle points 
of the lower momenta. Statistically significant differences between our 
results and PWA predictions are found. These differences vary with 
momentum and angle. The biggest 
difference, both in statistical significance and in size, is in the 
backward-angle region for momentum 139~MeV/$c$ where it is about 10\%. 
The size of the difference and its momentum and angle dependence are 
very unlikely to be explained by a single error in analysis, if such 
existed. In particular, note that the beam normalization, usually the 
strongest source of uncertainty, is very well controlled in this experiment.

\begin{figure}[p]
 \includegraphics[width=14cm]{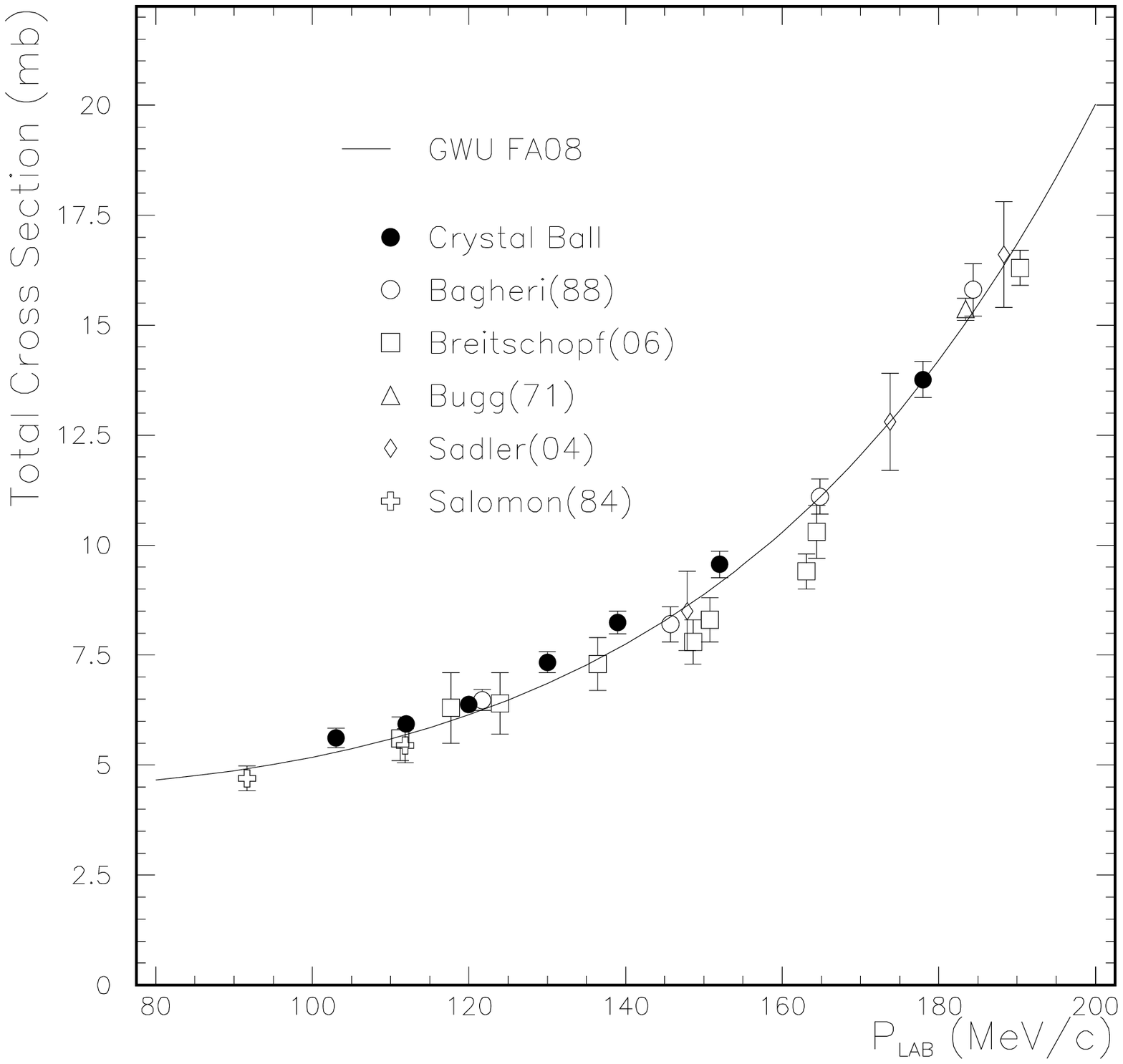}
 \caption{The total charge-exchange cross sections obtained from integrating 
the differential cross sections. The error bars show combined statistical 
and systematic uncertainties. The results are compared to the GWU FA08 
partial-wave analysis \cite{arndt06} and to previous data 
\cite{bagheri88,breitschopf06,bugg71,sadler04,salomon84}.}
 \label{fig:tot_cex}
 \end{figure}

The differential cross sections were fitted with an expansion in Legendre 
polynomials
\begin{equation} \label{eq:legendre}
\frac{d\sigma}{d\Omega} = \sum_{L\!=\!0}^{L_{\text{max}}} A_L\,P_L(\cos\theta)
\end{equation} 
to obtain the total cross sections, where $\sigma_T = 4\pi\,A_0$. 
Only sums up to $L_{\textrm{max}} = 2$ were needed for all momenta. 
The total charge-exchange cross sections are listed in Table~\ref{tab:tot_cex} together with statistical and total uncertainties. 
The Legendre coefficients and their uncertainties are listed in 
Table \ref{tab:legendre}.  The data are fitted very well within their 
uncertainties by Eq.\ \ref{eq:legendre}.
 \begin{table}[tbh!]
 \caption{The total charge-exchange cross section obtained from integrating 
the differential cross section with statistical and total uncertainties.}
 \label{tab:tot_cex}
 \begin{ruledtabular}
 \begin{tabular}{cccc}
Momentum & Total cross  & Statistical & Total \\
 (MeV/$c$) & section (mb) & uncertainty & uncertainty \\
\hline
103 & 5.61 & 0.10 & 0.22 \\
112 & 5.96 & 0.09 & 0.18 \\
120 & 6.39 & 0.09 & 0.19 \\
130 & 7.33 & 0.11 & 0.24 \\
139 & 8.21 & 0.09 & 0.26 \\
152 & 9.58 & 0.10 & 0.30 \\
178 & 13.75 & 0.11 & 0.41
 \end{tabular}
 \end{ruledtabular}
 \end{table}
 \begin{table}[tbh!]
 \caption{Legendre coefficients, with uncertainties in parentheses, 
from fits to the differential cross section data.}
 \label{tab:legendre}
 \begin{ruledtabular}
 \begin{tabular}{cccc}
Momentum & & & \\
(MeV/$c$) & $A_0$ & $A_1$ & $A_2$ \\
\hline
103 & 0.4463 (0.0078)  &  -0.540 (0.015)  &  0.110 (0.014) \\
112 & 0.4747 (0.0074)  &  -0.627 (0.014)  &  0.156 (0.013) \\
120 & 0.5081 (0.0071)  &  -0.695 (0.014)  &  0.192 (0.013) \\
130 & 0.5830 (0.0084)  &  -0.813 (0.016)  &  0.245 (0.015) \\
139 & 0.6532 (0.0070)  &  -0.946 (0.014)  &  0.346 (0.012) \\
152 & 0.7623 (0.0081)  &  -1.090 (0.016)  &  0.452 (0.015) \\
178 & 1.0939 (0.0089)  &  -1.361 (0.017)  &  0.729 (0.018)
 \end{tabular}
 \end{ruledtabular}
 \end{table}
The statistical and systematic uncertainties were added in quadrature for 
the total uncertainty. The results are shown in Fig.\ \ref{fig:tot_cex} and 
compared to the GWU FA08 partial-wave analysis and to the previous data 
\cite{bagheri88,breitschopf06,bugg71,sadler04,salomon84}. As expected from 
the differential cross sections, our results agree with PWA predictions for 
the momentum 178~MeV/$c$, and are higher then predictions for the other momenta.
\subsection{Consistency Checks \label{sec:consistency}}
\begin{figure}[p]
 \includegraphics[width=10cm]{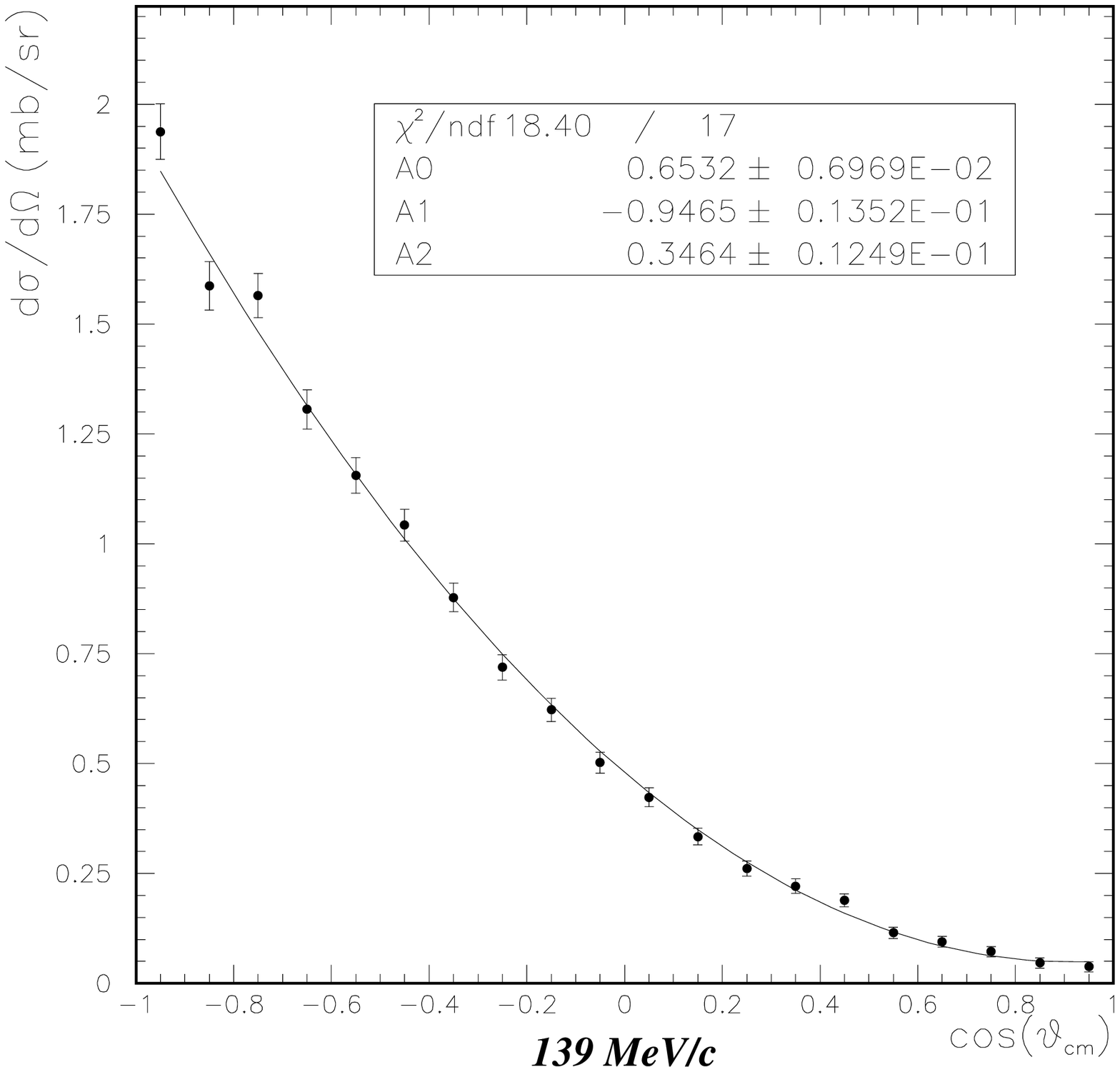}
 \caption{Fit of the differential cross section for 139~MeV/$c$ with an 
expansion in Legendre polynomials up through $L = 2$. Values of the 
$\chi^{2}$ of the fit and Legendre coefficients $A_{0}$, $A_{1}$, 
and $A_{2}$ are shown.}
 \label{fig:leg_fit_139}
 \end{figure}

\begin{figure}[p]
 \includegraphics[angle=90.,width=10cm]{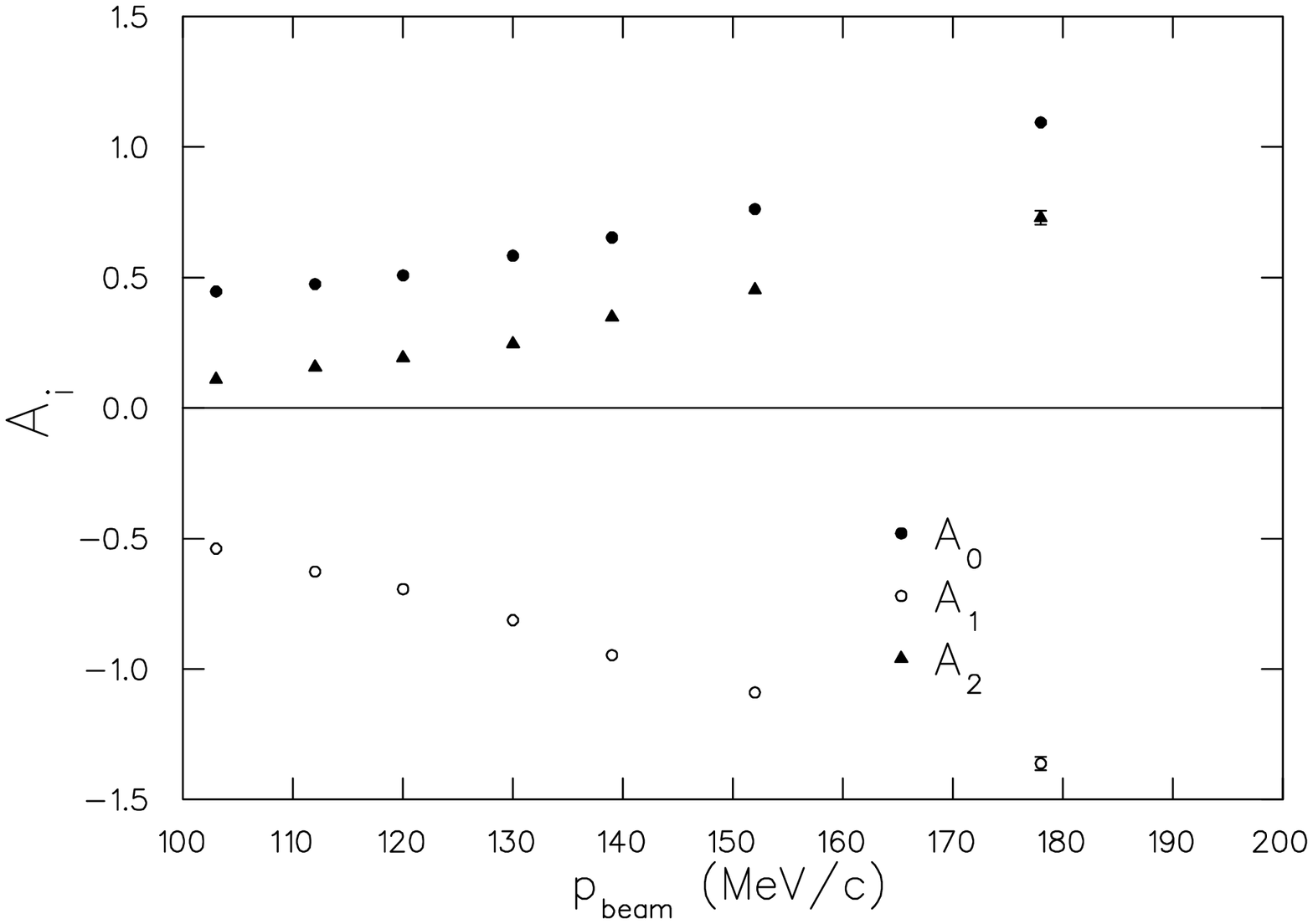}
 \caption{Legendre coefficients from fits to the differential cross 
section data.}
 \label{fig:a_coefs}
 \end{figure}

The results of the analysis were carefully examined to identify any issues 
that might not already be covered by the known systematic uncertainties. 
The first one has already been mentioned.  Although the GWU partial-wave 
predictions systematically disagree with the data at the backward angles, 
as seen in Figs.\ \ref{fig:dcs_4} to \ref{fig:dcs_1}, the Legendre 
expansion provides an excellent representation at all momenta with no 
need for terms beyond $P_2(\cos\theta)$, see Fig.\ \ref{fig:leg_fit_139} for example.
 Furthermore, as shown in Fig.\ \ref{fig:a_coefs}, the Legendre coefficients each have a smooth dependence 
on momentum.  These data can also be very well represented within the 
error bars by quadratic or cubic polynomial fits.  Such fits, however, 
are not reliable much outside the momentum range of the data.  Partial-wave codes are needed for such extrapolations.

The smooth behavior of the Legendre coefficients gives confidence that 
there are no significant momentum-dependent concerns in the data.  The 
$A_0$ coefficient provides values for the total cross sections.  As seen 
in Fig.\ \ref{fig:tot_cex}, the results of our experiment are consistent 
with those of other experiments, and generally have smaller uncertainties.

A final consistency check is the location of the minimum in the $0^{\circ}$ 
cross section, which arises from the interference of $s$ and $p$ waves. 
Although our error bars for these cross sections are larger than those 
of Jia {\it{et al.}} \cite{jia08}, we find the minimum cross section 
to be at $41.8 \pm 1.1$~MeV (116.0~MeV/$c$) as compared to $41.9 \pm 0.9$~MeV
 for Jia {\it{et al.}}.  The GWU FA08 solution yields $\simeq 46$~MeV.
\section{Conclusion}
Differential cross sections of the charge-exchange reaction 
$\pi^{-}p\rightarrow \pi^{0}n$ are presented for seven momenta in the range 
from 103 to 178~MeV/$c$. Complete angular distributions were obtained by using 
the Crystal Ball detector. The results presented here almost double the 
existing database in the low-energy region. The obtained cross sections 
are higher in the backward-angle region than the predictions of the GWU FA08 
partial-wave analysis based on earlier experiments. These data could have 
an important impact on investigations of isospin breaking at low energies. 
They will also be useful for extracting important physical quantities such 
as the $\pi$N $\sigma$ term.
\\
\begin{acknowledgments}
This work was supported in part by the U.S. DOE and NSF, by the Croatian 
MZOS, by the Russian Foundation for Basic Research, and by NSERC of Canada.
The assistance of BNL and AGS with the setup is greatly appreciated.
\end{acknowledgments}
\bibliography{cex_paper_a}
\end{document}